\documentclass[PRD,twocolumn,showpacs,superscriptaddress,preprintnumbers,nofootinbib,amsmath,amssymb]{revtex4-1}

\usepackage{xcolor}
\usepackage{graphicx}
\usepackage{dcolumn}
\usepackage{bm}
\usepackage{multirow}
\usepackage{ulem}

\usepackage{epstopdf}

\def\Mpl{M_{\mathrm{ Pl}}}

\newcommand{\beq}{\begin{equation}}
\newcommand{\eeq}{\end{equation}}
\newcommand{\beqa}{\begin{eqnarray}}
\newcommand{\eeqa}{\end{eqnarray}}

\newcommand{\bk}{{\mathbf{k}}}
\newcommand{\bx}{{\mathbf{x}}}

\newcommand{\bq}{{\mathbf{q}}}

\newcommand{\nn}{\nonumber}

\def\TT{{\rm TT}}
\newcommand{\half}{{\frac{1}{2}}}

\newcommand{\mn}{{\mu\nu}}

\newcommand{\hI}{\hspace{1cm}}

\newcommand{\hV}{\hspace{.5cm}}

\def\de{\delta}

\def\la{\lambda}

\def\De{\Delta}
\def\Ga{\Gamma}

\def\La{\Lambda}

\newcommand{\be}{\begin{equation}}
\newcommand{\ee}{\end{equation}}
\newcommand{\bea}{\begin{eqnarray}}
\newcommand{\eea}{\end{eqnarray}}
\newcommand{\ba}{\begin{array}}
\newcommand{\ea}{\end{array}}
\newcommand{\bit}{\begin{itemize}}
\newcommand{\eit}{\end{itemize}}
\newcommand{\vev}[1]{\left\langle#1\right\rangle}

\newcommand{\lan}{\left\langle}
\newcommand{\ran}{\right\rangle}

\newcommand{\GW}{{_{\rm GW}}}

\newcommand{\rhogw}{\rho_{\text{GW}}}
\newcommand{\Omgw}{\Omega_{\text{GW}}}

\newcommand{\SpecDen}[1]{P_{#1}}

\newcommand{\uetcT}{\mathcal{U}}

\newcommand{\uetc}{UETC}
\newcommand{\uetcs}{UETCs}
\newcommand{\eom}{\textit{eom}}
\newcommand{\ie}{\textit{i.e.}}
\newcommand{\eg}{\textit{e.g.}}

\definecolor{olivegreen}{rgb}{0,0.6,0}

\begin{document}

\title{Irreducible background of gravitational waves from a cosmic defect network:\\ update and comparison of numerical techniques}

\author{Daniel G. Figueroa}
\email{daniel.figueroa@ific.uv.es}
\affiliation{Instituto de F\'isica Corpuscular (IFIC),  University of Valencia-CSIC, E-46980, Valencia, Spain}
\author{Mark Hindmarsh}
\email{mark.hindmarsh@helsinki.fi}
\affiliation{Physics Department, University of Helsinki and Helsinki Institute of Physics, P. O. Box 64, FI-00014 Helsinki, Finland}
\affiliation{Department of Physics \& Astronomy, University of Sussex, Brighton, BN1 9QH, United Kingdom}
\author{Joanes Lizarraga}
\email{joanes.lizarraga@ehu.eus}
\affiliation{Department of Theoretical Physics, University of the Basque Country UPV-EHU, 48040 Bilbao, Spain}
\author{Jon Urrestilla}
\email{jon.urrestilla@ehu.eus}
\affiliation{Department of Theoretical Physics, University of the Basque Country UPV-EHU, 48040 Bilbao, Spain}

\date{\today}

\begin{abstract}
Cosmological phase transitions in the early Universe may produce relics in the form of a network of cosmic defects. Independently of the order of a phase transition, topology of the defects, and their global or gauge nature, the defects are expected to emit gravitational waves (GWs) as the network energy-momentum tensor adapts itself to maintaining {scaling}. We show that the evolution of any defect network (and for that matter any scaling source) emits a GW background with spectrum $\Omega_{\rm GW} \propto f^3$ for $f \ll f_0$, $\Omega_{\rm GW} \propto 1/f^2$ for $f_0 \lesssim f \lesssim f_{\rm eq}$, and $\Omega_{\rm GW} \propto~const$ (i.e.~exactly scale-invariant) for $f \gg f_{\rm eq}$, where $f_0$ and $ f_{\rm eq}$ denote respectively the frequencies corresponding to the present and matter-radiation equality horizons. This background represents an irreducible emission of GWs from any scaling network of cosmic defects, with its amplitude characterized only by the symmetry breaking scale and the nature of the defects. Using classical lattice simulations 
we calculate the GW signal emitted by defects created after the breaking of a global symmetry $O(N) \rightarrow O(N-1)$. 
We obtain the GW spectrum for $N$ between 2 and 20 
with two different techniques: integrating over unequal time correlators of the energy momentum tensor, 
updating our previous work on smaller lattices, 
and for the first time, comparing the result with the real time evolution of the tensor perturbations sourced by the same defects. Our results validate the equivalence of the two techniques. Using CMB upper bounds on the defects' energy scale, we discuss the difficulty of detecting this GW background in the case of global defects.
\end{abstract}


\maketitle

\section{Introduction}\label{sec:Intro}

The direct detection~\cite{Abbott:2016blz,Abbott:2016nmj,Abbott:2017vtc,Abbott:2017gyy,Abbott:2017oio} of gravitational waves (GWs) by advanced LIGO~\cite{Harry:2010zz} and advanced VIRGO~\cite{Acernese:2015gua}, represent a milestone in astronomy, and have opened a new window for exploring the Universe. Other GW detectors have just started observation (KAGRA~\cite{Somiya:2011np}), or have been approved (LIGO-India~\cite{Unnikrishnan:2013qwa,LISA-India}). A next generation of detectors is already being planned, including the {Einstein Telescope}~\cite{Sathyaprakash:2012jk} on the ground, and the {Laser Interferometer Space Antenna} (LISA) \cite{Audley:2017drz} and Taiji \cite{Hu:2017mde,Guo:2018npi} in space. Other proposals for space-based detectors, include DECIGO ({Deci-hertz Interferometer Gravitational wave Observatory}) \cite{Seto:2001qf,Kawamura:2006up}, and 
BBO ({Big Bang Observatory})~\cite{Crowder:2005nr}. Sufficiently energetic processes in the early Universe leave behind characteristic signatures in stochastic GW backgrounds, 
which are beginning to be constrained from the ground~\cite{LIGOScientific:2019vic,Renzini:2019vmt}. Space-based detectors will place more stringent constraints on early Universe scenarios,  
and have greater potential to detect cosmological sources~\cite{Caprini:2018mtu,Caprini:2019pxz}. Gravitational waves are, in fact, the most promising cosmic relic to probe many of the currently unknown details of the early Universe. It is important therefore to characterize all possible stochastic backgrounds in order to achieve a better understanding of a future detection. 

Whenever there is  an energy-momentum tensor with a non-zero transverse-traceless (\TT) part, GWs are expected to be emitted. In the absence of any source, GWs are also generated quantum mechanically during inflation, with an almost scale-invariant spectrum~\cite{Grishchuk:1974ny,Starobinsky:1979ty, Rubakov:1982df,Fabbri:1983us}. Depending on the underlying high energy physics modeling of inflation, active sources may also be present, generating GWs with a large amplitude and blue tilt~\cite{Anber:2006xt,Sorbo:2011rz,Pajer:2013fsa,Namba:2015gja,Ferreira:2015omg,Peloso:2016gqs,Domcke:2016bkh,Bartolo:2016ami,Thorne:2017jft}. After inflation, a kination dominated phase may also induce a  large blue tilt in the inflationary GW backgrond~\cite{Giovannini:1998bp,Giovannini:1999bh,Boyle:2007zx,Figueroa:2018twl,Figueroa:2019paj,Bernal:2019lpc}, whereas non-equilibrium phenomena can lead to a strong production of GWs, from particle production at (p)reheating~\cite{Easther:2006gt,GarciaBellido:2007dg,GarciaBellido:2007af,Dufaux:2007pt,Dufaux:2008dn,Dufaux:2010cf,Enqvist:2012im,Figueroa:2013vif,Figueroa:2014aya,Figueroa:2016ojl,Figueroa:2017vfa,Adshead:2018doq,Adshead:2019igv,Adshead:2019lbr} and oscillon dynamics~\cite{Zhou:2013tsa,Antusch:2016con,Antusch:2017vga,Liu:2017hua,Amin:2018xfe}, to strong first order phase transitions~\cite{Kosowsky:1992rz,Kamionkowski:1993fg,Caprini:2007xq,Huber:2008hg,Caprini:2009fx,Caprini:2009yp,Hindmarsh:2013xza,Hindmarsh:2015qta,Caprini:2015zlo,Hindmarsh:2017gnf,Cutting:2018tjt, Jinno:2019bxw} and cosmic string networks~\cite{Vachaspati:1984gt,Damour:2000wa,Damour:2001bk,Damour:2004kw,Sanidas:2012ee,Sousa:2013aaa,Blanco-Pillado:2017oxo,Auclair:2019wcv}. For a review on early Universe GW cosmological backgrounds, see~\cite{Caprini:2018mtu}.

In this paper we study the GW background sourced by a self-similar energy-momentum tensor perturbation.
Self-similarity, or scaling, means that the length scale of the energy-momentum distribution is proportional to the cosmic time. 
Scaling is exhibited by cosmic defects~\cite{Vilenkin:2000jqa,Hindmarsh:1994re}, which are a natural by-product of a phase transition in the early Universe. Cosmic defects also create  anisotropies in the Cosmic Microwave Background (CMB)~\cite{Ade:2013xla,Lizarraga:2014xza,Charnock:2016nzm,Lizarraga:2016onn,Lopez-Eiguren:2017dmc}. The search for cosmic defects in the CMB corresponds precisely to studying the imprint of the metric perturbations created by the defects. The B-mode polarization signal in the CMB is partly created by the tensor metric perturbation that will form the GW background we study in this work. 

Cosmic defects also exhibit other potentially observable effects: non-Gaussianity in the CMB~\cite{Figueroa:2010zx,Ringeval:2010ca,Regan:2014vha}, lensing events \cite{Vilenkin:1984ea,Bloomfield:2013jka}, and cosmic rays from the decay of strings into particle radiation~\cite{Brandenberger:1986vj,Srednicki:1986xg,Bhattacharjee:1991zm,Damour:1996pv,Wichoski:1998kh,Peloso:2002rx,Sabancilar:2009sq,Vachaspati:2009kq,Long:2014mxa,Auclair:2019jip}. In the case of Nambu-Goto strings, a stochastic background of GWs is expected from the oscillations of the resulting loops, chopped off from the main string network through cosmic history~\cite{Vilenkin:1981bx,Vachaspati:1984gt,Accetta:1988bg,Caldwell:1991jj,Siemens:2006yp,DePies:2007bm,Olmez:2010bi,Sousa:2013aaa,Sousa:2014gka,Blanco-Pillado:2017oxo,Ringeval:2017eww,Auclair:2019wcv}. The search for this background places the most stringent bounds on the energy scale of Nambu-Goto strings, see e.g.~\cite{Sanidas:2012ee,Blanco-Pillado:2017rnf,Ringeval:2017eww,Abbott:2017mem,LIGOScientific:2019vic}.  

As mentioned, cosmic defects are formed in early Universe phase transitions, which are usually described as a spontaneous symmetry-breaking process, driven by some scalar field(s) acquiring a non-zero expectation value within a vacuum manifold $\mathcal{M}$. If the manifold is topologically non-trivial, \ie~has a non-trivial homotopy group $\pi_n(\mathcal{M}) \neq \mathcal{I}$, topologically non-trivial field configurations will arise, producing strings for $n=1$, monopoles for $n=2$, and textures for $n=3$~\cite{Kibble:1976sj}. For higher $n$, the symmetry-breaking field is not topologically obstructed from reaching the vacuum manifold at any point in space-time, and non-topological field configurations arise. 
In cases when the symmetry broken is global, all non-constant field configurations produce energy-momentum, and are loosely  referred to  as global defects. When the broken symmetry is gauged and the non-trivial homotopy groups have $n$ = 0,1 or 2,  local defects appear. Cosmic strings, whether global or gauged, as well as any type of global defect exhibit scaling behavior, sufficiently long after the completion of the phase transition that created them~\cite{Turok:1991qq,Vilenkin:2000jqa,Hindmarsh:1994re,Durrer:2001cg}. All cases, topological or not, local or global, will be referred to as cosmic defects. 

In a previous letter~\cite{Figueroa:2012kw}, which will be referred to as {Paper~I} from now on, we clarified the origin of the scale-invariance of the GW background emitted by the self-ordering process of non-topological textures, arising after a global phase transition~\cite{Krauss:1991qu,JonesSmith:2007ne,Fenu:2009qf,Giblin:2011yh}. We generalized further the result, showing that any scaling source at the era of radiation domination (RD)  produces a GW background with a scale-invariant energy density power spectrum. In the case of cosmic defects, we emphasized that this is not related to their particular topology, or to the order of the phase transition, or to the global or local nature of the symmetry-breaking process that generated them. It is just a consequence of scaling and being in  RD. Using lattice simulations as an input, in Paper~I we also calculated numerically the GW amplitude from a system of global O($N$) defects, providing evidence that the numerical result converges to the analytical result calculated in the large-$N$ limit in~\cite{Fenu:2009qf,DaniPhD}.

While in Paper~I we clarified the origin of the scale-invariance of the GW spectrum and determined how the GW signal approaches (as we increase $N$) the analytical large-$N$ approximate result, various pertinent questions remain yet to be answered: how does the GW spectrum change when the GWs are emitted during matter domination (MD)? What is the GW spectrum when the defects arise from the breaking of a gauge symmetry, as in the case of local strings? Can the GW background leave an observable imprint in the cosmic microwave background? Can it be detected with pulsar timing arrays (PTA)? and with direct detection GW interferometers? Given the potential relevance of a detection of this background, it seems particularly pertinent to improve the details on the prediction of the signal itself. 

In the present paper, we update and complement the results from Paper~I. 
Firstly,  
we extend our prediction to a broader range of frequencies, studying the GWs emitted by a scaling source during both RD and MD. This introduces a new feature in the spectrum, which does not remain scale-invariant within the entire frequency range. We study also the GW spectrum at super-horizon scales. In particular, we find that the energy density power spectrum scales as $h^2\Omega_{\rm GW} \propto f^3$ for $f \ll f_{0}$, where $f_0$ is the frequency today corresponding to the present horizon. The spectrum reaches a maximum at $f = f_0$, and between $f = f_0$ and $f = f_{\rm eq}$ scales as $h^2\Omega_{\rm GW} \propto 1/f^2$, with $ f_{\rm eq}$ the frequency today corresponding to the horizon at the moment of matter-radiation equality. Eventually, for $f \gg f_{\rm eq}$, the spectrum settles down to a scale-invariant amplitude $h^2\Omega_{\GW} \propto const.$, as reported in~Paper~I. 

Secondly, we update the numerical input used in~Paper~I based on the extraction of the unequal-time (\uetc) correlators of the transverse-traceless part of the energy-momentum tensor  from field theory lattice simulations; namely we use new simulations with a larger volume, from which we obtain {\uetc}s with a wider spectral range. We also present the reconstruction of the GW spectrum based on the sum over a weighted eigenvalue-eigenvector decomposition coming from the diagonalization of the \uetc. 

Thirdly, and most importantly, 
we present a complementary numerical calculation of the gravitational wave energy density power spectrum, 
which allows two non-trivial checks of our results:\vspace*{0.1cm}

$\bullet$  We obtain the GW spectrum by following the real time evolution of the tensor metric perturbations, as they are continuously sourced by the defect network itself. We compare for the first time the GW spectra from the \uetc\ method with those produced by real time evolution from the same energy-momentum source. We discuss the circumstances under which a good agreement is found within an appropriate spectral range. The success of this comparison provides a validation of both methods, suggesting that the use of either method should be equally acceptable in future numerical GW computations.

 \vspace*{0.1cm}

$\bullet$ The prediction of the GW power spectra from defect networks (say scale-invariant in RD) is rooted on the assumption of perfect scaling. We have checked that by `switching on' the defect source term in the equations of motion of the tensor perturbations before the defect network has reached the scaling regime, this does not produce the predicted spectrum from scaling as it should. The reason behind this is that in such cases the spectrum reflects the highly random initial field configuration, which eventually prevents the signal from forming the expected scaling profile. This result highlights the importance of initiating the GW evolution only when the network is in scaling.\vspace*{0.1cm}

Finally we discuss the amplitude today of the GW backgrounds from different defect networks, based on our simulations. We compare our results with previous studies of GW production from global defects available in the literature.

We note that the case $N = 2$, corresponding to global strings, is anomalous, in that our updated GW power is of order a factor $\sim 2$ bigger than the value given in Paper I. We argue this is connected with the special difficulties in assessing the scaling of the network, as discussed in \cite{Hindmarsh:2019csc}. This case is specially relevant as it can be connected e.g.~with string-inspired models~\cite{Dasgupta:2004dw,Burgess:2008ri} that enjoy (approximate) global symmetries with low $N$, as well as with axion-like dark matter candidates, where a network of global strings is naturally expected to be produced, see \eg\ \cite{Marsh:2015xka}. Further work is required in order to make a robust prediction of its GW signal. 

The paper is divided as follows. In Sect.~\ref{sec:GW} we  review the basic aspects of stochastic GW backgrounds with sources. In Sect.~\ref{sec:GWfromCDmethods} we turn our attention to GWs sourced by scaling seeds. We derive previous and new aspects of the frequency dependence of the GW spectrum, depending on the cosmic epoch. For comparison, we review briefly the analytical calculation of the GW spectrum in RD, in the case of a global phase transition $O(N) \rightarrow O(N-1)$ with $N\gg 1$ (further details are shown in Appendix~\ref{app:1}). In Sect.~\ref{sec:GWfromUETC} we discuss some aspects of the methodology of our lattice simulations, and we present the definition and extraction of the tensor unequal-time correlator (\uetc) from them. We then give numerical examples of the GW energy density spectra obtained with this technique, using our new numerical simulations of global defects, based on a $O(N) \rightarrow O(N-1)$ symmetry breaking with arbitrary $N$. We also show the reconstruction of the GW spectrum through partial summation of weighted terms obtained from the diagonalization of the tensor \uetc. In Sect.~\ref{sec:GWfromRealTimeEvol} we present our numerical results for the GW energy density spectrum from the same simulations introduced in  Sect.~\ref{sec:GWfromUETC}, but obtained from the real time evolution (in real space) of the tensor perturbations, while being sourced by the defect network. We discuss the limitations to reconstruct the spectrum by this method, and the circumstances required to reach a good agreement with the method based on \uetc 's. In Sect.~\ref{sec:Conclusions} we summarize our results, highlight some of the technical difficulties involved in our numerical calculations, and discuss the difficulty to detect this GW background. 

\begin{center}
------------------
\end{center}

We work in $\hbar = c = 1$ units, with $M_{\rm Pl} = 1/\sqrt{G} \approx 1.22\times10^{19}\,{\rm GeV}$ the Planck mass, and $G$ Newton's constant. Summation over repeated indices is assumed.

\section{Gravitational waves}
\label{sec:GW}

We will study the evolution of the fields and the gravitational waves, when the Universe is well described by a spatially flat Friedman-Lema\^itre-Robertson-Walker (FLRW) metric, sourced by a perfect fluid. 
Including the relevant metric perturbation, the line element is written as
\begin{equation}
ds^{2}=a^{2}(t)\left[-dt^{2}+\left(\delta_{ij}+h_{ij}\right)dx^{i}dx^{j}\right],\label{eq:metric}
\end{equation}
with $a(t)$ the scale factor, $t$ conformal time, and the metric perturbations $h_{ij}$ are transverse ($\partial_{i}h_{ij} = 0$) and traceless ($h_{i}^{i} = 0$). 

Splitting the Einstein equations into background and linearized equations, the GW equations of motion (\eom) in a FLRW background are (see e.g.~\cite{Caprini:2018mtu})
\begin{equation}\label{eq:GWeomBasic}
\ddot{\bar{h}}_{ij}\left(\mathbf{x},t\right)-\left(\nabla^{2}+\frac{\ddot{a}(t)}{a(t)}\right)\bar{h}_{ij}\left(\mathbf{x},t\right)=16\pi Ga(t)\Pi_{ij}^{\rm TT}\left(\mathbf{x},t\right),
\end{equation}
where we have introduced a conformal redefinition of the tensor perturbations $\bar{h}_{ij}\left(\mathbf{x},t\right)=a(t)h_{ij}\left(\mathbf{x},t\right)$, and dots denote derivatives with respect to the conformal time. The source $\Pi_{ij}^{\rm TT}$ is the TT-part of the anisotropic stress tensor $\Pi_{ij}$, which we define below. The conditions $\partial_i \Pi_{ij}^{\rm TT} = \Pi_{ii}^{\rm TT} = 0$ hold for $\forall\,\bx, \forall\,t$. Either in RD or in MD, or in general for a scale factor with a power law behaviour in time, it holds that $\ddot{a}/{a} \sim \mathcal{H}^{2}$, where $\mathcal{H} \equiv \dot a/a$ is the (comoving) Hubble rate. Hence, the term $\ddot{a}/{a}$ is negligible at sub-horizon scales $k \gg \mathcal{H}$, and therefore we will drop it from now on. The \eom\ of sub-horizon modes in Fourier space can then be written as
\begin{equation}\label{eq:h_e.o.m.}
\ddot{\bar{h}}_{ij}\left(\mathbf{k},t\right)+k^{2}\bar{h}_{ij}\left(\mathbf{k},t\right)=16\pi G\,a(t)\Pi_{ij}^{\rm TT}\left(\mathbf{k},t\right)\,,
\end{equation}
where $\bk$ is the comoving wave-number and $k=|\mathbf{k}|$ its modulus. The solution to Eq.~(\ref{eq:h_e.o.m.}) is given by a convolution with the Green's function associated to a free wave-operator in Minkowski spacetime, $G_{>}(k,t-t') = k^{-1}\sin[k(t-t')]$. That is, at times $t > t_I$, with $t_{I}$ an initial time with no gravitational waves, $h_{ij}\left(\mathbf{k},t_{I}\right) = \dot{h}_{ij}\left(\mathbf{k},t_{I}\right) = 0$, we obtain
\begin{eqnarray}\label{eq:h(Pi)}
h_{ij}(k,t) &=& {\bar{h}_{ij}(k,t)\over a(t)} \\
&=& \frac{16 \pi G}{k\,a(t)}\int_{t_{I}}^t dt' a(t')\sin[k(t-t')]\Pi_{ij}^{\TT}(k,t')\,.\nn
\end{eqnarray}

Obtaining the TT-part of a tensor in configuration space amounts to a non-local operation. It is more convenient to do it in Fourier space, where a projector filtering out only the TT degrees of freedom  of a tensor can be easily written down. The GW source can then be written as
\begin{equation}
\Pi_{ij}^{\rm TT}(\mathbf{k},t) = \Lambda_{ij,lm}(\hat\bk)\,\Pi_{lm}(\mathbf{k},t),
\end{equation}
where $\Lambda_{ij,lm}(\hat\bk)$ is a projection operator defined as
\begin{eqnarray}\label{projector}
\indent \Lambda_{ij,lm}(\mathbf{\hat k}) \equiv P_{il}(\hat\bk)
P_{jm}(\hat\bk) - {1\over2} P_{ij}(\hat\bk)
P_{lm}(\hat\bk),\,\\
P_{ij} = \delta_{ij} - \hat k_i \hat k_j\,,\hV \hat k_i = k_i/k\hI\hV \,.
\label{lambda}
\end{eqnarray}
Thanks to the fact that $P_{ij}\hat k_j = 0$ and $P_{ij}P_{jm} = P_{im}$, one can easily see that the transverse-traceless conditions in Fourier space, $k_i\Pi_{ij}^{\rm TT}(\hat\bk,t) = \Pi_{ii}^{\rm TT}(\hat\bk,t) = 0$, are satisfied at any time. 

The anisotropic stress tensor $\Pi_{\mn}$ describes the deviation of an energy momentum tensor $T_{\mn}$ with respect to a  perfect fluid. The spatial-spatial components read
\begin{equation}
\Pi_{ij} \equiv T_{ij} - p\,g_{ij}\,,
\end{equation}
with $p$ the homogeneous background pressure and $g_{ij}=a^2(t)(\delta_{ij}+h_{ij})$ the spatial-spatial FLRW perturbed metric. In the scenarios we consider in this paper the energy density is dominated by a homogeneous and isotropic perfect fluid. The energy-momentum of this background has spatial-spatial components $T^{\rm pf}_{ij} = p\,g_{ij}$, with the pressure either one third of the energy density (RD) or zero (MD).

On top of  this there is a sub-dominant contribution from cosmic defects, which have their own energy-momentum tensor $T_{ij}^{\rm def}$. Hence, in these scenarios, the (spatial-spatial components of the) total energy-momentum tensor are given by $T_{ij} = T^{\rm pf}_{ij} + T^{\rm def}_{ij}$. It is clear then that $\Pi_{ij} = T_{ij}^{\rm def}$, so that the active source of GWs in our case is the TT-part of the cosmic defects' energy-momentum tensor.

\subsection{Spectrum of gravitational waves}

Expanding the Einstein equations to second order in the tensor perturbations, one recognizes that the energy density of a GW background is given by~\cite{MaggioreBookI}
\begin{eqnarray}\label{eq:GWrhoCont}
\rho_{\GW}(t) &=& \frac{1}{32\pi G a^2(t)}\left\langle \dot{h}_{ij}(\bx,t)\dot{h}_{ij}(\bx,t)\right\rangle_V \\
&\equiv & \frac{1}{32\pi G a^2(t)}\frac{1}{V}\int_V d\bx\, \dot h_{ij}(\bx,t)\dot h_{ij}(\bx,t) \nn\\
&=& \frac{1}{32\pi G a^2(t)}\int\frac{d\bk}{(2\pi)^3}\frac{d\bk'}{(2\pi)^3}~\dot h_{ij}(\bk,t)\dot h_{ij}^*(\bk',t)\nn\\
&& \hspace*{2cm}\times\,\frac{1}{V}\int_V\hspace*{-1mm} d\bx~e^{-i\bx(\bk-\bk')}\nn\,,
\end{eqnarray} 
with $\langle...\rangle_V$ a spatial average over a sufficiently large comoving volume $V$ encompassing all the relevant wavelengths of the $h_{ij}$ perturbations. In the limit $kV^{1/3} \gg1 $, $\int_{_{\rm V}}\hspace*{-1mm} d\bx~e^{-i\bx(\bk-\bk')}$ $\rightarrow$ $(2\pi)^3\delta^{(3)}(\bk-\bk')$, and hence
\begin{equation}
 \rho_{\GW}(t) = \frac{1}{32\pi G a^2(t) V}\int\frac{d\bk}{(2\pi)^3}~\dot h_{ij}(\bk,t)\dot h_{ij}^*(\bk,t)\,.
\end{equation}
The GW energy density spectrum per logarithmic interval is defined as
\bea 
\rho_{\GW}(t) \equiv \int\frac{d\rho_{\GW}}{d\log k}\,d\log k\,,\hspace* {1.5cm}\\
\frac{d\rho_{\GW}}{d\log k} = 
\frac{k^3}{(4\pi)^3 G\,a^2(t)V} \int \frac{d\Omega_k}{4\pi}\,\dot h_{ij}(\bk,t)\dot h_{ij}^*(\bk,t)\,, 
\label{eq:GWrhoContSpectrum}
\end{eqnarray}
where $d\Omega_k$ represents a solid angle element in $\bk$-space. 

In our case, GWs are created from a network of  cosmic defects. As the symmetry breaking process that originates the defects is a random process, we cannot predict the exact location of each cosmic defect. However, we can still describe the stochastic distribution that characterizes the defect network. The spatial distribution of the GW will therefore be assumed to be also stochastic, following the random distribution of the defects. Applying the {ergodic hypothesis}, we can replace $\langle...\rangle_V$ by an ensemble average $\langle...\rangle$ over realizations. The stochastic background of GWs can then be described by
\begin{eqnarray}\label{eq:GW_energyDensity}
\rho_{\mathrm{\GW}} &=& \frac{1}{32\pi G a^2(t)}\left\langle \dot{h}_{ij}(\bx,t)\dot{h}_{ij}(\bx,t)\right\rangle \nn\\
&=& \frac{1}{32\pi G a^2(t)}\int\frac{d\mathbf{k}}{\left(2\pi\right)^{3}}\frac{d\mathbf{k}'}{\left(2\pi\right)^{3}}~e^{i\bx(\mathbf{k-k'})}\nn\\
&& \hspace{2cm}\times\left\langle\dot{{h}}_{ij}\left(\mathbf{k},t\right)\dot{{h}}_{ij}^{*}\left(\mathbf{k'},t\right)\right\rangle \,.
\end{eqnarray}
The expectation value in the second line of Eq.~(\ref{eq:GW_energyDensity}), assuming statistical homogeneity and isotropy, 
can be written as
\begin{equation}
\left\langle\dot{{h}}_{ij}\left(\mathbf{k},t\right)\dot{{h}}_{ij}^{*}\left(\mathbf{k'},t\right)\right\rangle \equiv (2\pi)^3\,\SpecDen{\dot h}(k,t)\,\delta^{(3)}(\bk-\bk')\,,
\end{equation}
so that we can write
\begin{equation}\label{eq:GWspectByPhdot}
\rho_{\mathrm{\GW}}(t) = \frac{1}{(4\pi)^3 G a^2(t)}\int\hspace*{0mm} {dk\over k}~k^3\,\SpecDen{\dot h}(k,t)  \,.
\end{equation}
From here we define the {\it GW energy density power spectrum} as
\begin{eqnarray}\label{eq:GWrhoStochaSpectrum}
\frac{d\rho_{\GW}}{d\log k}(k,t) = \frac{1}{(4\pi)^3 G\,a^2(t)}\,k^3\,\SpecDen{\dot h}(k,t)\,,
\end{eqnarray}
which will be referred to as the {GW power spectrum} (or simply as the {GW spectrum}). 

Obtaining $\SpecDen{\dot h}(k,t)$ can be done with the help of Eq.~(\ref{eq:h(Pi)}), by first writing
\begin{equation}
\dot{h}_{ij}(\mathbf{k},t) = \frac{16\pi G}{ka(t)}\int_{t_{I}}^{t}dt'a(t')\,\mathcal{G}(k(t-t'))\,\Pi_{ij}^\TT(\mathbf{k},t'),\label{eq:Dh(Pi)}
\end{equation}
with $\mathcal{G}(k(t-t')) \equiv \left(k\cos[k(t-t')]-\mathcal{H}\sin[k(t-t')]\right)$. This leads to
\begin{eqnarray}\label{eq:P_doth_expanded}
&& \SpecDen{\dot h}(k,t) = \frac{(16\pi G)^2}{k^2a^2(t)}\int_{t_{I}}^{t}dt'\int_{t_{I}}^{t}dt''a(t')a(t'')\\
&& \hspace{2cm}\times\,\mathcal{G}(k(t-t'))\,\mathcal{G}(k(t-t''))\,\Pi^2(k,t',t'')\,,\nn
\end{eqnarray} 
where we have introduced the unequal time correlator (\uetc) of the TT-part of the anisotropic-stress $\Pi_{ij}^{\TT}$, 
\begin{eqnarray}\label{eq:UETC}
\left\langle {\Pi}_{ij}^\TT(\bk,t)\,{{\Pi}_{ij}^{\TT}}(\bk',t')\right\rangle \equiv (2\pi)^3\,{\Pi}^2(k,t,t')\,\delta^{(3)}(\bk-\bk')\,.\nn\\
\end{eqnarray}

Once GW production ends, GWs propagate as free waves, each mode oscillating with period $T_k = {2\pi/ k}$. We need therefore to take a time average over the product of $\mathcal{G}(\bk,t,t')$ functions, 
\begin{eqnarray}\label{eq:GreenFuncOscAv}
\lan\mathcal{G}(\bk,t,t')\mathcal{G}(\bk,t,t'')\ran_{T_k} &\equiv& {1\over T_k}\int_t^{t+T_k}\hspace*{-0.5cm}d\tilde t~\mathcal{G}(\bk,\tilde t,t')\mathcal{G}(\bk,\tilde t,t'')\nn\\ 
&=& {1\over2}(k^2+\mathcal{H}^2(t))\cos[k(t'-t'')]\nn\,.\\
\end{eqnarray}
Replacing $\mathcal{G}(\bk,t,t')\,\mathcal{G}(\bk,t,t'')$ by $\lan\mathcal{G}(\bk,t,t')\mathcal{G}(\bk,t,t'')\ran_{T_k}$ in Eq.~(\ref{eq:P_doth_expanded}), and taking into account that at subhorizon scales $(k^2+\mathcal{H}^2(t)) \approx k^2$, we arrive at
\begin{eqnarray}\label{eq:PShdot}
\SpecDen{\dot h} &=& \frac{(16\pi G)^2}{2a^2(t)}\int_{t_{I}}^{t}dt'\int_{t_{I}}^{t}dt''a(t')a(t'')\\
&& \hspace*{1.5cm}\times\,\cos[k(t'-t'')]\,\Pi^2(k,t',t'')\nn\,.
\end{eqnarray}
Plugging Eq.~(\ref{eq:PShdot}) into~Eq.~(\ref{eq:GWrhoStochaSpectrum}), we finally find the GW energy density power spectrum of a stochastic background of GW (at subhorizon scales) as
\begin{eqnarray}\label{eq:GW_spectra(Pi)}
\frac{d\rho_{\mathrm{\GW}}}{d\log k}\left(k,t\right) &=& \frac{2}{\pi}\,{G\,k^3\over a^4(t)} \int_{t_{I}}^{t}dt'\int_{t_{I}}^{t}dt''\,a(t')\,a(t'')\\
&& \hspace*{1.5cm}\times\,\cos[k(t'-t'')]\,\Pi^2(k,t',t'')\,.\nn
\end{eqnarray}
For convenience, we can also normalize the GW energy density spectrum to the critical density $\rho_c \equiv 3H^2 / 8\pi G$, obtaining
\begin{eqnarray}\label{eq:OmGW_spec}
 \Omega_{\rm GW}(k,t) &\equiv & {1\over\rho_c}\frac{d\rhogw}{d\log k} \\ &=& {16\,G^2\,k^3\over 3H^2a^4(t)}\int_{t_{I}}^{t}dt'\int_{t_{I}}^{t}dt''\,a(t')\,a(t'')\nn \\
&& ~~~~~~~~~~~~~~\times \cos[k(t'-t'')]\,\Pi^2(k,t',t'')\,.\nn
\end{eqnarray}

\section{Gravitational waves from Scaling Seeds}\label{sec:GWfromCDmethods}

Based on causality and dimensional grounds, Ref.~\cite{Krauss:1991qu} originally argued that the field dynamics following after a global phase transition should generate an approximately scale-invariant background of GWs. The amplitude of such background was estimated with the GW quadrupole approximation, without any reference to the number of components $N$ of the corresponding symmetry breaking field. 

In the context of a phase transition driven by the breaking of a global O($N$) symmetry into a O($N-1$) group, even though the field equations are non-linear, analytic calculations can be carried out in the $N \gg 1$ limit, describing the evolution of the non-topological global defects that emerge after the phase transition, see Ref.~\cite{Turok:1991qq}. Within such context and using a full treatment of the tensor metric perturbation (i.e.~without resorting to the quadrupole approximation), Refs.~\cite{JonesSmith:2007ne,Fenu:2009qf} demonstrated that in the large $N$ limit, an exact scale-invariant background of GWs is generated (during RD) by the self-ordering dynamics of the non-topological global defects arising after the O($N$) $\longrightarrow$ O($N-1$) symmetry breaking. On the numerical side, Ref.~\cite{Giblin:2011yh} studied lattice simulations after a second-order phase transition, concluding that the global defects created in that case generate a GW background consistent with scale invariance, even though the numerical spectra exhibited some tilt and oscillatory fluctuations. 

In this section we will generalize the above results, deriving and discussing the common aspects of the spectral shape of the stochastic GW background emitted by any network of cosmic defects in a scaling regime (which will be simply referred to as {scaling seeds}). In particular, in Sect.~\ref{subsec:GWsubH}, we first review our findings from Paper I during RD, and then we extend the results to the production of GWs from the evolution of scaling seeds during MD. In Sect.~\ref{subsec:GWsuperH} we characterize the GW background at super-horizon scales for both RD and MD. In Sect.~\ref{subsec:GWtoday} we discuss the overall spectral shape of the GW background spanned over all frequencies, and in particular the form of the resulting red-shifted spectrum today. Finally, in~\ref{subsec:GWanalytics}, we review the analytic estimation of the GW signal emitted by self-ordering scalar fields based on the large $N$ limit of global defects. We postpone the presentation of our numerical results from lattice simulations to Sections~\ref{sec:GWfromUETC} and \ref{sec:GWfromRealTimeEvol},  where two different numerical methods for obtaining the spectrum of GWs emitted by a network of cosmic defects will be presented. There we will also compare the analytic formulation of this section with the outcome from the numerical simulations.

\subsection{GW spectrum at sub-horizon scales.}
\label{subsec:GWsubH}

The origin of the scale-invariance of the GW background emitted by the self-ordering process of non-topological defect was 
clarified in Paper I. There it was demonstrated that any scaling source with a non-vanishing transverse-traceless energy momentum tensor always produce a background of GWs during RD, with an exact scale-invariant energy density power spectrum. As emphasized in Paper I, the result is just a consequence of the defects' scaling behaviour and of being in RD.

Let us recall that once we know the \uetc\ of the tensor anisotropic stress $\Pi^2(k,t_1,t_2)$ (\ref{eq:UETC}), we can compute the spectrum of GWs emitted by simply plugging $\Pi^2(k,t_1,t_2)$ into Eq.~(\ref{eq:GW_spectra(Pi)}). In the case of a defect network, the correlator $\Pi^2(k,t_1,t_2)$ can be obtained, from field theory simulations~\cite{Bevis:2006mj,Lizarraga:2012mq,Daverio:2015nva,Hindmarsh:2016lhy,Lopez-Eiguren:2017dmc,Hindmarsh:2018wkp}. In the specific case of non-topological defects arising after spontaneous symmetry breaking of a global O($N$), $\Pi^2(k,t_1,t_2)$ can be also estimated analytically in the large $N$ limit, as we will review in~\ref{subsec:GWanalytics}. 

Before we consider the explicit form of the \uetc\ from a defect network, let us recall the most fundamental property of any network of cosmic defects: whenever cosmic defects are created during a phase transition, the resulting defect network (after the phase transition is completed) enters gradually into a scaling regime, where the number density of defects per comoving Hubble volume $\sim 1/\mathcal{H}^3 \sim t^3$, remains invariant through cosmic history~\cite{Kibble:1976sj,Hindmarsh:1994re,Vilenkin:2000jqa}. Once in the scaling regime, the \uetc\ can only depend on $k$ through the dimensioness variables $x_1 = kt_1$ and $x_2=kt_2$. From dimensional analysis it is forced to take the form
\begin{equation}\label{eq:UETCscaling}
\Pi^2(k,t_1,t_2) = {4v^4\over\sqrt{t_1t_2}}\,\uetcT(kt_1,kt_2),
\end{equation}
where $v$ is the vacuum expectation value (VEV) in the broken state of the scalar fields (the factor 4 is a convention to match  the tensor UETC of Ref.~\cite{Bevis:2010gj}). 

Using the scaling form of the correlator, 
\begin{eqnarray}\label{eq:GWspectrum3}
 \Omega_{\rm GW}(k,t) &=& {64k^2\over 3H^2a^4(t)}\left({v\over M_{\rm Pl}}\right)^4\\
&\times& \int dx_1 dx_2 \,{a_1a_2\over\sqrt{x_1x_2}}\,\cos(x_1-x_2)\,\uetcT(x_1,x_2)\nn\,,
\end{eqnarray}
where $a_1 \equiv a(x_1/k)$, $a_2 \equiv a(x_2/k)$. 

Before the late time accelerated expansion of the Universe, 
but after the era of electron-positron annihilation, 
the scale factor can be written as 
\bea\label{eq:aMIXED} 
a(t) &=& a_{\rm eq}\left([(\sqrt{2}-1)(t/t_{\rm eq})+1]^2-1\right)\nn\\
&=& a_0^3\Omega_{\rm mat}^{(0)}{H_0^2t^2\over 4} + a_0^2\sqrt{\Omega_{\rm rad}^{(0)}}H_0t\,,
\eea
where $a_{\rm eq}$ is the scale factor at the time of matter-radiation equality, $t_{\rm eq}$. 
In the second expression we have used the integral representation 
\bea\label{eq:tEQintegral} 
a_0H_0t_{\rm eq} &=& \int_{z_{\rm eq}}^\infty {dz\over\sqrt{\Omega_{\rm rad}^{(0)}(1+z)^4 + \Omega_{\rm mat}^{(0)}(1+z)^3}}\nn\\
&=& {2(\sqrt{2}-1)\over(1+z_{\rm eq})}{1\over \sqrt{\Omega_{\rm rad}^{(0)}}}\,.
\eea
Changes of the number of relativistic degrees of freedom ($dof$) during RD can be easily taken into account by correcting the solution deep in the radiation era as
\begin{eqnarray}
a_{\rm RD}(t) = a_0^2\sqrt{\Omega_{\rm rad}^{(0)}}H_0\int \mathcal{R}_{t'}dt'\,,
\end{eqnarray}
where
\begin{eqnarray}
\mathcal{R}_{t} \equiv \left({g_{s,0}\over g_{s,t}}\right)^{4/3}\left({g_{{\rm th},t}\over g_{{\rm th},0}}\right)\, ,
\end{eqnarray}
with $g_{s,t}$ and $g_{{\rm th},t}$ the entropic and thermal energy density number of relativistic $dof$ at time $t$. 
For most of cosmic history, $g_{s,t} \simeq g_{{\rm th},t}$, and it is a good approximation to treat $\mathcal{R}$ 
as a piecewise constant function. We take $\mathcal{R}_{t} \simeq \mathcal{R}_{\rm QCD} \simeq 0.39$ before the quark-gluon QCD phase transition $t < t_{\rm QCD}$, $\mathcal{R}_{t} \simeq \mathcal{R}_{\rm e^-e^+} \simeq 0.81$ between QCD and electron-position annihilation $t_{\rm QCD} < t < t_{\rm e^-e^+}$, and $\mathcal{R}_{t} \simeq \mathcal{R}_{0} = 1$ after electron-position annihilation $t > t_{\rm e^-e^+}$. The scale factor Eq.~(\ref{eq:aMIXED}) during RD can then be approximated as 
$$a_{\rm RD}(t) \simeq \sqrt{\Omega_{\rm rad}^{(0)}}a_0^2H_0\mathcal{R}_{*}t\,,$$
with $\mathcal{R}_{*} \simeq \mathcal{R}_{\rm QCD}, ~\mathcal{R}_{\rm e^-e^+}$ or $\mathcal{R}_{0}$, depending on the time $t$. 

Plugging this behavior into Eq.~(\ref{eq:GWspectrum3}), leads to a sub-horizon spectrum of GW, for modes that become sub-horizon $x \equiv kt \gg 1$ during RD, as
\begin{eqnarray}\label{eq:GWspectrum2}
\Omega_{\rm GW}(x,t) = \Omega_{\rm rad}(t)\left({v\over\Mpl}\right)^{\hspace*{-1mm}4}{\hspace*{-0.5mm}}\,{\mathcal{R}_*\over \mathcal{R}_t} {F}_{\rm RD}^{[\mathcal{U}]}(x)\,,\hspace*{0.8cm}\\
\label{eq:F_U}
{F}_{\rm RD}^{[\mathcal{U}]}(x) \equiv {64\over3}\int^x\hspace*{-0.3cm} dx_1 \int^x \hspace*{-0.3cm} dx_2~ \sqrt{x_1x_2} \cos(x_1-x_2)\,\uetcT(x_1,x_2)\nn\,,\nn
\end{eqnarray} 
where $\Omega_{\rm rad}(t) = 1$ while $t \ll t_{\rm eq}$, and $\Omega_{\rm rad}(t) < 1$ for $t > t_{\rm eq}$. At subhorizon scales, $\uetcT(x_1,x_2)$ is peaked near $x_1 = x_2 \equiv x$, and decays along the diagonal as a power law $\propto x^{-p}$, with $p$ a positive real number, see \eg~\cite{Durrer:2001cg}. Hence the convergence of the integration is guaranteed as long as the decay is fast enough, \ie~$p > 2$. In such a case ${F}_{\rm RD}^{[\mathcal{U}]}(x)$ becomes more and more insensitive to its upper bound of integration, approaching asymptotically a constant value for $x \gg 1$. In other words $F_{\uetcT}(x \gg 1)$ approaches the constant ${F}_{\rm RD}^{(\infty)} \equiv {F}_{\rm RD}^{[\mathcal{U}]}(x\rightarrow \infty)$. As a consequence of this, the GW spectrum at subhorizon wavelengths becomes scale-invariant.

For every type of defect there is a characteristic function $\uetcT(x_1,x_2)$, and thus a well determined value ${F}_{\rm RD}^{(\infty)}$, which gives the magnitude of the GW spectrum. Its value today, with the traditional reference value for the Hubble rate $H_0 = 100h_0\,{\rm km}\,{\rm s}^{-1}\, {\rm Mpc}^{-1}$, is 
\begin{eqnarray}\label{eq:GWspectrumToday}
h_0^2\Omgw^{(0)}(k)  = h_0^2\Omega_{\rm rad}^{(0)}\left({v\over\Mpl}\right)^{\hspace*{-1mm}4}{\hspace*{-0.5mm}}\mathcal{R}_*{F}_{\rm RD}^{(\infty)}.
\end{eqnarray}
The GW background produced during RD by the evolution of any network of defects in scaling regime is therefore exactly scale-invariant, modulo a mild step-wise change in $\mathcal{R}_*$ due to the evolution of the number of relativistic species in thermal equilibrium. The amplitude of the GW background today is suppressed by the fraction of relativistic species $\Omega_{\rm rad}^{(0)}$, and is proportional to the fourth power the VEV as $(v/\Mpl)^4$. It also depends on the shape of the \uetc ~(and hence on the type of defect), which ultimately modulates the amplitude through $F_{\rm RD}^{(\infty)}$. 

Eq.~(\ref{eq:GWspectrumToday}) summarizes the theoretical results from Paper I about GW emission from RD, incorporating now as a new element the change in the number of relativistic species. We generalize next the analysis to the emission of GWs during MD. For times when the radiation component is completely sub-dominant, $t \gg t_{\rm eq}$, the scale factor Eq.~(\ref{eq:aMIXED}) can be written approximately as $a(t) \simeq {1\over4}a^3_0\Omega_{\rm mat}^{(0)}H_0^2t^2$. As soon as the \uetc\ during MD is scaling, it can be written again as in Eq.~(\ref{eq:UETCscaling}). For simplicity we assume scaling is maintained for $t \geq t_{\rm eq}$. Using Eq.~(\ref{eq:GWspectrum3}), together with Eq.~(\ref{eq:tEQintegral}), the spectrum of GW at sub-horizon scales $x \equiv kt \gg 1$ during MD reads
\begin{eqnarray}\label{eq:GWspectrum4}
\Omega_{\rm GW}(x,t) = \Omega_{\rm rad}(t)\left({v\over\Mpl}\right)^{\hspace*{-1mm}4}\,{k_{\rm eq}^2\over k^2}\,{F}_{\rm MD}^{[\mathcal{U}]}(x)\,,\nonumber\\
\end{eqnarray}
where $x_{\rm eq} \equiv kt_{\rm eq}$, $k_{\rm eq} \equiv {1/ 2t_{\rm eq}}$, and 
\begin{eqnarray}
\label{eq:F_MD}
&& {F}_{\rm MD}^{[\mathcal{U}]}(x) \equiv {64\over3}(\sqrt{2}-1)^2\\
&& ~~~~ \times \int_{x_{\rm eq}}^x\hspace*{-0.3cm} dx_1 \int_{x_{\rm eq}}^x \hspace*{-0.3cm} dx_2~ (x_1x_2)^{3/2}\cos(x_1-x_2)\,\uetcT(x_1,x_2)\,.\nn
\end{eqnarray} 
By construction $k_{\rm eq}$ corresponds to the mode with half wavelength ${\pi/ k_{\rm eq}}$ equal to the horizon $1/t_{\rm eq}$ at the time of matter-radiation equality. We can redshift this spectrum today as
\begin{eqnarray}\label{eq:GWspectrumTodayMD}
h_0^2\Omgw^{(0)}(k) &\equiv& {1\over\rho_{\rm c}}\left(\frac{d\rhogw}{d\log k}\right)\nn\\
&=& h_0^2\Omega_{\rm rad}^{(0)}\left({v\over\Mpl}\right)^{\hspace*{-1mm}4}{\hspace*{-0.5mm}}\left(k_{\rm eq}\over k\right)^2{F}_{\rm MD}^{(\infty)}\,.
\end{eqnarray}

\subsection{GW spectrum at super-horizon scales.}
\label{subsec:GWsuperH}

Let us note that Eq.~(\ref{eq:GWspectrumToday}) and Eq.~(\ref{eq:GWspectrumTodayMD}) correspond only to the GW energy density spectrum for modes well inside the current horizon, during RD and MD, respectively. The energy density carried by GWs is only meaningful for sub-Hubble modes, 
because the notion of an energy-momentum tensor associated to GWs requires that the wavelengths of the GWs are much smaller than the characteristic length scale of the background metric (see e.g.~\cite{MaggioreBookI}). In the case of the FLRW background, a characteristic length scale at any moment is the causal horizon, which is given (modulo factors of order unity) by the instantaneous Hubble radius. The universe is however expected to be homogeneous and isotropic beyond the causal horizon, so in principle we can extend formally the notion of GW energy density spectrum at super-Hubble scales $x \equiv kt \ll 1$. The time scale for the oscillation of a given super-horizon mode is however much larger than the age of the universe, $T_k = 2\pi/k = (2\pi/x)t\gg t$, so there is no sense in averaging over oscillations as in Eq.~(\ref{eq:GreenFuncOscAv}). Instead, our starting point must be Eq.~(\ref{eq:P_doth_expanded}) with the functions $\mathcal{G}(k,t,t')$, $\mathcal{G}(k,t,t'')$ evaluated in the super-Hubble limit $kt,kt',kt'' \ll 1$. 

In order to proceed we note that 
the Green's function (\ref{eq:h(Pi)}) is only valid for sub-Hubble modes $x \gg 1$. Hence, we must derive the general Green's function for the full differential equation (\ref{eq:GWeomBasic}).
For power-law expansion rate $a(t) \propto t^p$, we find
\begin{eqnarray}
\label{eq:GeneralGreenFct}
 h_{ij}(k,t) &=& {16\pi G \over k a(t)}\int_{t_I}^t dt' a(t')\,G^{(p)}_{>}(x,x')\,\Pi_{ij}^{\TT}(k,t')
 \label{eq:GeneralGreenFctII}
\end{eqnarray}
where
\begin{eqnarray}
G^{(p)}_{>}(x,x') &\equiv&  x x'\left\lbrace j_{p-1}(x')y_{p-1}(x) - j_{p-1}(x)y_{p-1}(x')\right\rbrace,\nn\\
\end{eqnarray}
and  $j_p(x), y_p(x)$ are spherical Bessel functions of the first and second kind. 
In terms of the functions $G^{(p)}_>$, the Green's functiond for $\dot{h}$ are
\begin{equation}
\mathcal{G}^{(p)}(k,t,t') = a(t)\frac{d}{dt}\frac{G^{(p)}_{>}(x,x')}{a(t)} 
\end{equation}
For RD ($p=1$) we have 
$$
G^{(1)}_{>}(x,x') = \sin(x-x').
$$
while for MD ($p=2$) we obtain  
$$
G^{(2)}_{>}(x,x') = {1\over x x'}[(1 + x x')\sin(x - x')- (x - x') \cos(x - x')].
$$ 
The subhorizon Green's function used in Eq.~(\ref{eq:h(Pi)}) therefore can be applied at arbitrary scales in RD, 
which follows from $a''(t) = 0$, but only at large $x$ and $x'$ in MD.

We are now ready to obtain at super-horizon scales, for arbitrary power law expansion rate $a(t) \propto t^p$. 
We find $ \mathcal{G}(k,t,t') \longrightarrow  k(1-\mathcal{H}(t-t'))$, where $\mathcal{H} = p/t$. In particular, for RD   and MD   we have
\begin{eqnarray}\label{eq:SuperHorizonGreen}
\mathcal{G}(k,t,t') \xrightarrow[kt,kt' \ll 1]{} 
\left\lbrace\begin{array}{cl}
k({t'/t}) &\,,~{\rm RD}\vspace*{3mm}\\
k\left[-1+2({t'/t})\right] &\,,~{\rm MD}
\end{array}\right.
\end{eqnarray}
We can now consider Eq.~(\ref{eq:P_doth_expanded}) to represent a valid formal expression for the energy density spectrum of GWs at super-Hubble scales, as long as we use the expressions of $\mathcal{G}(k,t,t')$ given in Eq.~(\ref{eq:SuperHorizonGreen}). This implies that Eq.~(\ref{eq:GW_spectra(Pi)}) is also valid at super-horizon scales as long as we replace $\cos(k(t'-t''))$ by either expression from Eq.~(\ref{eq:SuperHorizonGreen}). The energy density spectrum for scaling seeds Eq.~(\ref{eq:GWspectrum3}) is then valid as well at super-Hubble scales, as long as we replace $\cos(x_1-x_2)$ by ${x_1x_2/ x^2}$ (for RD) or by $(-1+2x_1/x)(-1+2x_2/x)$ (for MD). Alternatively, identical replacements can be done in the specialized expressions for RD Eq.~(\ref{eq:GWspectrum2}) and MD Eq.~(\ref{eq:GWspectrum4}).

In order to find the scale-dependence of the spectrum at super-Hubble scales, the upper bound on the integration (\ref{eq:F_MD}) 
is taken $x \ll 1$. 
At super-horizon scales, the \uetc\ scales as $\mathcal{U}(x_1,x_2) \simeq \mathcal{U}_{SH} = const$ for $x_1 \simeq x_2$, and $\mathcal{U}(x_1,x_2) \ll \mathcal{U}_{SH}$ otherwise \cite{Durrer:1997ep}.
Then, when we perform the integrals in ${F}_{\rm RD}^{[\mathcal{U}]}(x)$ and ${F}_{\rm MD}^{[\mathcal{U}]}(x)$, we can take $\mathcal{U}(x_1,x_2)$ out of the integrand, and substitute it simply by a constant. 
We obtain therefore that the GW energy density spectrum at super-hubble scales during RD, scales as 
\begin{eqnarray}
 \Omega_{\rm GW}^{\rm [RD]}(x\ll1) &\propto& \left[\int^{x \ll 1} \hspace*{-0.5cm} dx' \sqrt{x'}\,\left({x'\over x}\right)\right]^2 \propto x^3\,,
\end{eqnarray}
whereas for MD, scales like
\begin{eqnarray}
 \Omega_{\rm GW}^{\rm [MD]}(x\ll1) &\propto&  \left[{k_{\rm eq}\over k}\int^{x \ll 1} \hspace*{-0.5cm} dx' {x'}^{3/2} \left(2{x'\over x}-1\right)\right]^2
 \\ &\propto& (t/t_{\rm eq})^2x^3\,.\nonumber
\end{eqnarray}

\subsection{GW spectrum at all scales (at RD and today).}
\label{subsec:GWtoday}

According to our previous discussion, the GW energy density spectrum can be formally extended to 
super-Hubble scales as functions 
going as $\Omega_{\rm GW}(x\ll1) \propto  x^3 \propto \left(k/\mathcal{H}\right)^3$, independently of the expansion rate $a(t) \propto t^p$. Hence, at a fixed time during RD, the overall spectrum scales as $\Omega_{\rm GW}(x\ll 1) \propto k^3$ for super-horizon modes, and turns  into a scale-invariant spectrum $\Omega_{\rm GW}(x\gg1) \propto k^0$  for the modes that have already crossed the horizon. Later on, at a given moment during MD, the spectrum scales $\Omega_{\rm GW}(x\ll1) \propto k^3$ for super-horizon modes, reaches a maximum at the horizon scale, decreases as $\Omega_{\rm GW}(1 \ll x \ll x_{\rm eq}) \propto k^{-2}$ for scales that crossed during MD, and eventually settles down to a scale-invariant amplitude $\Omega_{\rm GW}(x\gg x_{\rm eq}) \propto k^0$ for modes that crossed the horizon during RD. 

\begin{figure}[t]
\includegraphics[width=8.7cm,height=5cm,angle=0]{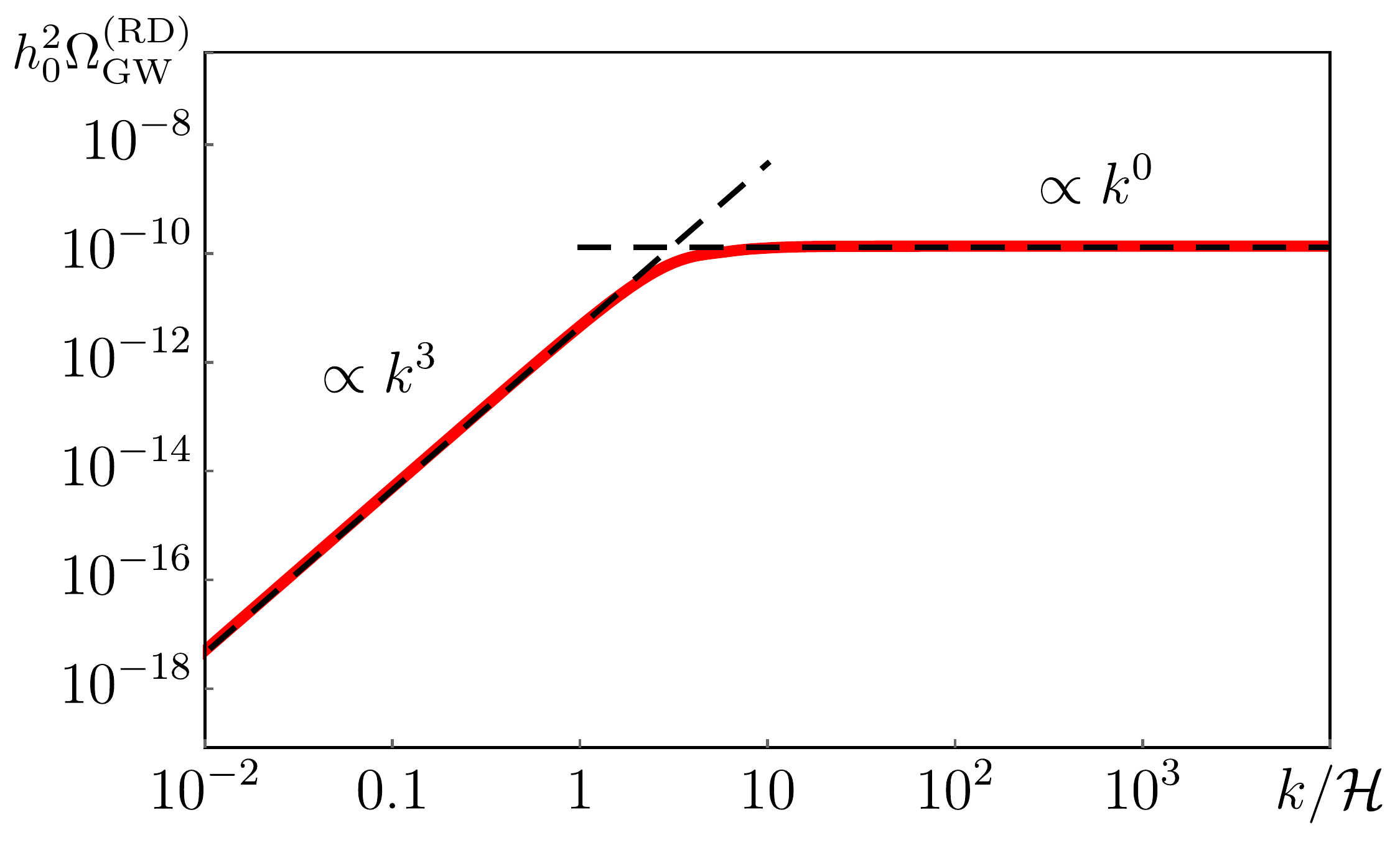}\vspace*{2mm}
\includegraphics[width=8.7cm,height=5cm,angle=0]{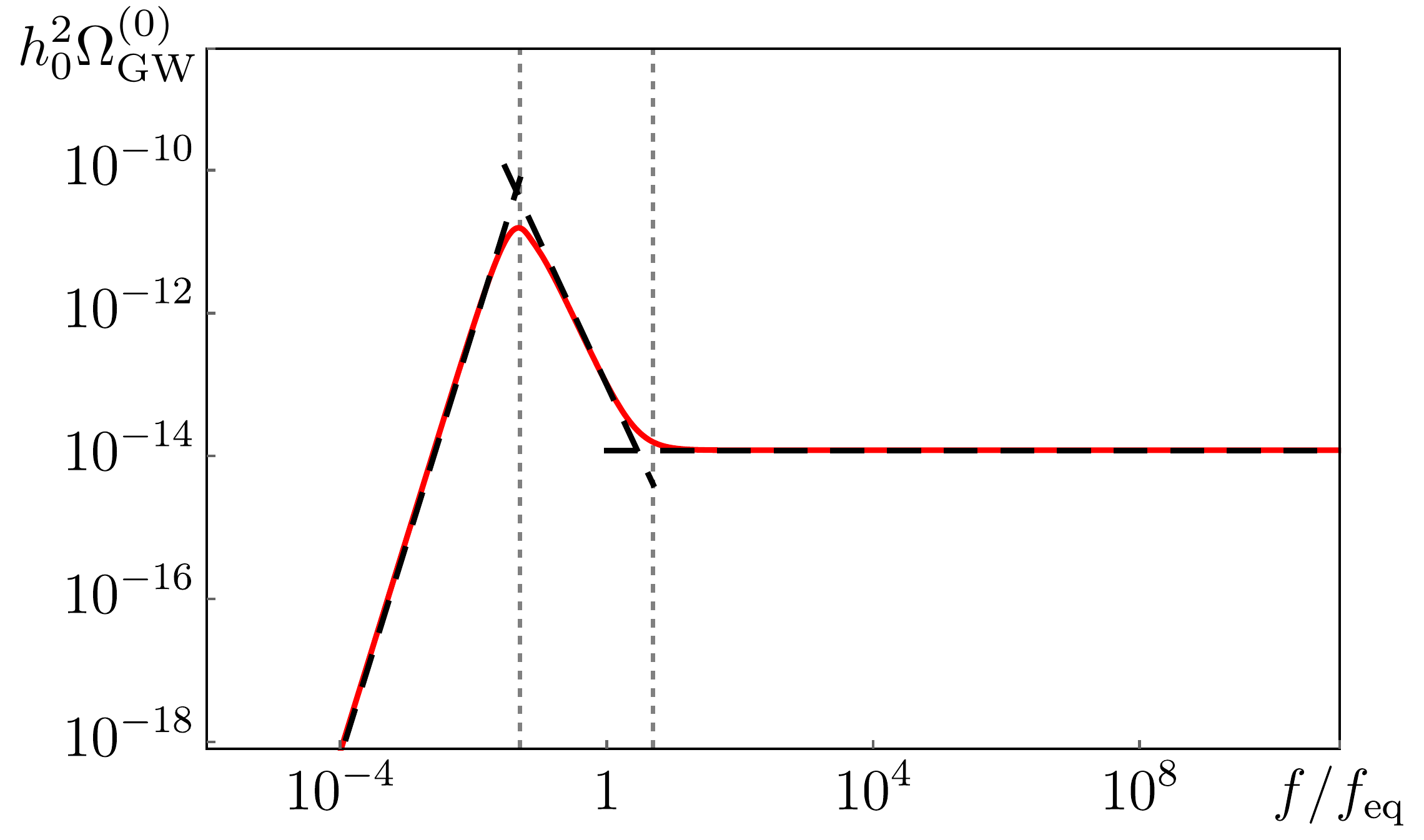}
\caption{{\it Top:} Instantaneous spectrum of GW during RD, $h_0^2\Omgw^{\rm (RD)}(k,t)$, plotted as a function of $x = k/\mathcal{H}$, assuming no change in the effective number of relativistic degrees of freedom between production and 
the time of evaluation (${\mathcal{R}_*/ \mathcal{R}_t} = 1$). 
{\it Bottom:} Red-shifted gravitational wave spectrum today $h_0^2\Omgw^{(0)}$ (using $\mathcal{R}_* = 1$). 
The dashed vertical lines represent, from left to right, the frequency $f_0$ of the present horizon (indicating the maximum of the spectrum), and the transition frequency $f_* = few \times f_{\rm eq}$ that signals the appearance of the high-frequency plateau.}
\label{fig:GWshapeDependence}
\end{figure}

In Fig.~\ref{fig:GWshapeDependence} we plot the full spectrum during RD [c.f.~Eq.~(\ref{eq:GWspectrum2})]
\begin{eqnarray}\label{eq:OmegaGW_duringRD}
\Omgw^{\rm (RD)}(k,t) = \Omega_{\rm rad}(t)\left({v\over\Mpl}\right)^{\hspace*{-1mm}4}{\mathcal{R}_*\over \mathcal{R}_t}{F}_{\rm RD}^{[\mathcal{U}]}(x)\,,
\end{eqnarray}
computed with the aid of {\uetc}s in scaling (for $N = 4$) that we obtain from lattice simulations in Sect.~\ref{sec:GWfromUETC}. We scale out ${\mathcal{R}_*/ \mathcal{R}_t}$ from the plot so that the steps due to the changing of the number of relativistic $dof$ are not shown, and hence we are left with exact power laws. By evaluating ${F}_{\rm RD}^{[\mathcal{U}]}(x)$ at all values $x = k/\mathcal{H}$, we obtain a continuous spectrum around horizon-crossing scales $x\sim 1$, smoothly interpolating the two asymptotic regimes ${F}_{\rm RD}^{[\mathcal{U}]}(x \gg 1) \longrightarrow {F}_{\rm RD}^{(\infty)}$ at sub-Hubble scales, and ${F}_{\rm RD}^{[\mathcal{U}]}(x \ll 1) \propto k^3$ at super-Hubble scales. Let us note that even though during RD, $\Omega_{\rm rad}(t) = 1$ holds by definition, it is convenient to maintain nonetheless such factor in Eq.~(\ref{eq:OmegaGW_duringRD}), as it controls the dilution of the GW spectrum once RD ends (see e.g.~Eq.~(\ref{eq:GWspectrumToday}), which describes the redshifted plateau amplitude today, suppressed by the current fraction of radiation to total energy density in the Universe $\Omega_{\rm rad}^{(0)} \ll 1$). 

Analogously, we plot in Fig.~\ref{fig:GWshapeDependence} the full redshifted spectrum at the present time (recall we are ignoring the effect of dark energy), by using the superposition of GW spectra obtained separately for RD and MD,
\begin{eqnarray}
h_0^2\Omgw^{(0)} &=& h_0^2\Omega_{\rm rad}^{(0)}\left({v\over\Mpl}\right)^{\hspace*{-1mm}4} \\
&& \times \left(\mathcal{R}_*{F}_{\rm RD}^{(\infty)}\Theta(x-x_{\rm eq}) + {k_{\rm eq}^2\over k^2}{F}_{\rm MD}^{[\mathcal{U}]}(x)\right)\,,\nonumber
\end{eqnarray}
so that by evaluating ${F}_{\rm MD}^{[\mathcal{U}]}(x)$ continuously at all values of $x$, we obtain a smooth spectrum around horizon-crossing scales $x\sim 1$, interpolating the super-Hubble regime ${F}_{\rm RD}^{[\mathcal{U}]}(x \ll 1) \propto k^3$ and the MD crossing scales regime  ${F}_{\rm RD}^{[\mathcal{U}]}(1 \ll x \ll x_{\rm eq}) \propto k^{-2}$, eventually matching smoothly around $x\sim x_{\rm eq}$ with the RD crossing scales regime ${F}_{\rm RD}^{[\mathcal{U}]}(x \gg x_{\rm eq}) \longrightarrow {F}_{\rm RD}^{(\infty)}$. We choose $\mathcal{R}_* = 1$ for simplicity in the bottom panel of Fig~\ref{fig:GWshapeDependence}.

Note that in obtaining the full GW spectrum today $h_0^2\Omgw^{(0)}$, we have approximated the transition from RD to MD at $t_{\rm eq}$ as instantaneous. 
The true interpolation of the spectrum between the regime $\Omega_{\GW} \propto k^{-2}$ for modes emitted during MD, and the regime $\Omega_{\GW} \propto const.$ for modes emitted during RD, would have small differences from that depicted in the bottom panel of Fig.~\ref{fig:GWshapeDependence}. However, as appreciated in the figure itself, the transition between these two regimes takes place at a frequency slightly larger\footnote{In Sect.~\ref{sec:GWfromRealTimeEvol} we will show that modes evolving during RD only form the plateau once they have became sufficiently small compared to the horizon scale. This explains why the transition scale between the $k^{-2}$ and $k^0$ regimes in the spectrum corresponds to a slightly shorter scale than the horizon at matter-radiation equality, i.e.~to a frequency $f_* = \alpha f_{\rm eq}$ with $\alpha \gtrsim 1$ a constant somewhat larger than unity.} than the frequency associated to the horizon scale at the moment of matter-radiation equality. The latter corresponds to a very large scale, and hence to a very small frequency today, which can be obtained as $f_{\rm eq} = {1\over 2\pi}{k_{\rm eq}\over a_0} $ = ${1\over 2}{a_{\rm eq}\over a_0}H_{\rm eq} \simeq$ ${1\over 2}(1+z_{\rm eq})^{1/2}\,H_0$, leading to
\begin{eqnarray}\label{eq:fEQ}
f_{\rm eq} \simeq 6.6\cdot 10^{-17} ~{\rm Hz}\,,
\end{eqnarray}
where we have used $k_{\rm eq} \equiv \pi a_{\rm eq} H_{\rm eq}$, $H_0/H_{\rm eq} \simeq (a_{\rm eq}/a_0)^{3/2}$, $a_0/a_{\rm eq} \simeq 1 + z_{\rm eq} \simeq 3400$ and $H_0 \simeq 70$ km/s/Mpc. Eq.~(\ref{eq:fEQ}) immediately informs us about an important aspect of the GW background we are studying: only the plateau part of the spectrum, emitted during RD, is relevant for direct observation.\footnote{This does not apply to the CMB, where the small frequency part $f \lesssim f_{\rm eq}$ of the GW spectrum can leave an imprint in the form of temperature and polarization anisotropies.} This is the case for pulsar timing array (PTA) experiments or current/planned direct detection interferometer experiments, as the typical frequencies accessible to the former are around $\sim 10^{-8}$ Hz, whereas the frequency range of the latter spans from $\sim 10^{-4}$ Hz to $\sim 10^{3}$ Hz (with huge frequency gaps in between). We discuss further the detectability of this GW background in Sect.~\ref{sec:Conclusions}.

\subsection{Analytical calculation of the GW background in the large-N limit of a global symmetry breaking}
\label{subsec:GWanalytics}

In a global theory where an $O(N)$ symmetry is spontaneously broken into $O(N-1)$, even though the field equations are nonlinear, analytic calculations are possible in the limit of large $N \gg 1$~\cite{Turok:1991qq,Boyanovsky:1999jg}. The starting point is an $N$-component scalar field $\Phi = (\phi_1,\phi_2,...,\phi_N)^{\rm T}/\sqrt{2}$ with lagrangian
\bea
-\mathcal{L} = (\partial_\mu\Phi)^{\rm T}(\partial^\mu\Phi) + \lambda\left(|\Phi|^2-v^2/2\right)^2 + \mathcal{L}_{\rm int}\,,
\label{Lagon}
\eea
where $\lambda$ is a dimensionless self-coupling, $v$ the VEV in the broken phase, $|\Phi|^2 = {1\over2}\sum_a\phi_a^2$, and $\mathcal{L}_{\rm int}$ represents some interaction with other degrees of freedom (\eg~a thermal bath or other scalar fields). When due to the dynamics (here unspecified) $\mathcal{L}_{\rm int}$ cannot compensate any further the tachyonic mass in the potential, the $O(N)$ symmetry is spontaneously broken to $O(N-1)$. As a result, $\Phi$ is driven 
to the vacuum manifold, given by $|\Phi({\bf x},t)|^2 = v^2/2$. Due to causality, in regions separated away by a comoving distance larger than the comoving horizon 
the values of $\Phi(\bx,t)$ and $\Phi(\bx',t)$ must be uncorrelated. As a consequence, gradient energy density is generated between disconnected regions. For $N \gg 1$, the dynamics of the Goldstone modes can be well described by a non-linear sigma model, where we force the vacuum constraint $\sum_a\phi_a^2(\bx,t) = v^2$ by a Lagrange multiplier. This approximation is very good for physical scales much larger than $m^{-1} \equiv 1/(\sqrt{\lambda}v)$. At large scales the field components are free to wander around in the vacuum manifold, giving rise to a gradient energy density which will generate GWs on those scales. 

Even though the \eom\ of the $N$ field components are non-linear, in practice the self-ordering dynamics of the fields can be described approximately by a linearized \eom\ when $N \gg 1$~\cite{Turok:1991qq}. A simple analytical solution of the field dynamics can be then found, showing explicitly that the self-ordering dynamics of the defects actually exhibit scaling.  Once the field solution is known, one can also calculate analytically, again in the large $N$-limit, the GW power spectrum emitted by the self-ordering dynamics of the non-topological defects. The full details of the calculation of the GW spectrum can be found in~\cite{Fenu:2009qf}, and for completeness we present a schematic derivation in Appendix~\ref{app:1}. Here we just quote the resulting GW spectral amplitude (today) for modes emitted during   RD~\cite{Fenu:2009qf,DaniPhD}
\bea\label{eq:GW_largeNfinal}
h_0^2\Omgw^{(0)}(f) \simeq \frac{650}{N}\, h_0^2\Omega^{(0)}_{\rm rad}\left(\frac{v}{{\Mpl}}\right)^{\!\!4}\hspace*{-0.1cm},
\eea
which, as expected for RD, corresponds to a scale invariant background (ignoring the change in the number of relativistic $dof$), with a spectral amplitude just characterized by a dimensionless number\footnote{While the original number found in~\cite{Fenu:2009qf} was $511/N$, after improving the numerical integration in~\cite{DaniPhD}, and testing the result against different accuracy levels and schemes of integration, it was found that a more correct number is $650/N$. The expressions given in either \cite{Fenu:2009qf} or \cite{DaniPhD} did not consider the change in the number of relativistic $dof$.}. There is no dependence either on the self-coupling $\lambda$ of the symmetry-breaking field, because the effective theory of the Goldstone modes, responsible for the creation of the GWs is a non-linear $\sigma$-model, and the coupling disappears when the scalar  mode is integrated out. Finally, the fact that the GW signal decreases with $N$  is also expected, as the larger the number of field components, the smaller the gradients between them, and hence the smaller the GWs emitted. We can identify the value of $F_{\rm RD}^{(\infty)}$ in the large $N$ analytical calculation Eq.~(\ref{eq:GW_largeNfinal}) as 
\begin{eqnarray}\label{eq:650overN}
F_{\rm RD}^{(\infty)}\big{|}_{N\to\infty} \simeq {650\over N}\,.
\end{eqnarray}
In section~\ref{sec:GWfromUETC} we will compare this number and its dependence on $N$ with the actual numerical values of $F_{\rm RD}^{(\infty)}$ calculated with the input of \uetc s obtained from numerical lattice simulations of the dynamics after the spontaneous global breaking of $O(N)$ into $O(N-1)$. As expected, the numerical results approach the analytical amplitude for $N \gg 1$, but disagree noticeably for small $N$.

From dimensional analysis it can also be deduced (using Eq.~(\ref{eq:GWlargeNanalytics})), that the GW spectrum for MD scales as $h_0^2\Omgw^{(0)}(f) \propto {f_{\rm eq}^2\over f^2}F_{\rm MD}^{(\infty)}\big{|}_{N\to\infty}$, with $F_{\rm MD}^{(\infty)}\big{|}_{N\to\infty}$ a constant that could be obtained from a numerical computation as we did for RD. Numerical integration (with sufficient accuracy) of the analytical spectrum is however costly (due to oscillations of the Bessel functions present in the solution to the self-ordering dynamics, see Appendix~\ref{app:1}). Furthermore, as mentioned before, from an observational point of view, only the part of the GW spectrum generated during RD is relevant, as only that part of the spectrum is potentially observable by GW experiments like PTAs, or present and planned direct detection GW interferometers. Since we have already clarified the overall shape of the GW spectrum over all frequencies, we will focus from now now, in the remaining of the paper, only in the RD case.

\section{Lattice computation of GWs. ~~~~~~~ Part 1:~Unequal-time-correlators}
\label{sec:GWfromUETC}

We present now our numerical results, based on lattice simulations. We use two different numerical methods to obtain the energy density spectrum of the GWs emitted by the network of cosmic defects. In this section, we discuss our results from lattice-based unequal time correlators (UETC) of the defects' energy-momentum tensor, which serve as an input to compute the GW spectrum. 

Our starting point for a lattice simulation of global defects is to consider a scalar field $\Phi$ with $N$ (real) components $\Phi = (\phi_1,\phi_2,...,\phi_N)^{\rm T}/\sqrt{2}$, and a potential $V(\Phi) =  \lambda\left(|\Phi|^2-v^2/2\right)^2$, where $\lambda$ is the dimensionless self-coupling, $v$ the vacuum expectation value (VEV) in the broken phase, and $|\Phi|^2 = {1\over 2}\sum_b\phi_b^2$. When due to some interaction with other degrees of freedom (represented either by another field or by a thermal plasma), the tachyonic mass in $V(\Phi)$ dominates over other interaction terms, the $O(N)$ symmetry is spontaneously broken to $O(N-1)$. The scalar field reaches a (spatially dependent) expectation value, very close to $|\Phi| =v/\sqrt{2}$ in most regions. 

For $N=2$ and $N =3$, the field remains zero along lines and points in space, creating strings and monopoles. 
The results for these cases are rather different to the ones obtained in the large $N$ limit considered before; both because 
of the presence of the topological defects, and because $N$ is not large, 
meaning that the non-linear terms in the field equations neglected in the large-$N$ approximation are important. 
As for higher $N$ no topological defects are formed, 
the numerical results should approach the analytic ones in large $N$ limit. 
The case $N=4$ is a boundary case. Here, the field can leave the vacuum manifold, but only at isolated spacetime points,  
in field configurations called textures. 
The difference between the linear and non-linear sigma model dynamics turns out to be minimal \cite{Bevis:2004wk,Urrestilla:2007sf}. 

In the following we explain the procedure to obtain the GW spectrum from a model with a global O($N$) symmetry, solving the \eom\ of the system and obtaining the unequal-time-correlators (\uetc) correspoding to the system. We simulate cases ranging from a small values of $N$, to cases closer and closer to the large-$N$ limit; in particular we consider $N=2,3,4,8,12,20$. The \eom\ for the field components are, in the continuum,
\bea
\ddot\phi_b + 2\mathcal{H}\dot\phi_b - \vec\nabla^2\phi_b =-2\lambda\, a^2\left(|\Phi|^2-v^2/2\right)\phi_b\,,
\label{eomuetc}
\eea
with $b = 1,\dots, N$, and where we have kept $\lambda$ as a parameter. The lattice version of this equation can be found in Appendix~\ref{app:2}. Here it is relevant to note that due to the presence of the factor $a^2$ on the $r.h.s.$ of equation (\ref{eomuetc}), the size of the defects, say the width of strings or the radius of monopoles, {shrink} in comoving coordinates, as the system evolves. This is a well-known and well-studied issue in lattice simulations of defect networks. As in previous studies in the literature, we will use the Press-Ryden-Spergel (PRS) method~\cite{Press:1989yh} to deal with this problem. 
This amounts to keeping  $a^2\la = \la_c = const.$, so that the physical scalar mass parameter shrinks as $m_\Phi = \sqrt{\la}v\propto 1/a$, and hence the width of the topological defects grows linearly with the scale factor $\delta l \sim 1/m_{\Phi} \propto a(t)$.\\

The algorithm we use in our numerical simulations, solves the lattice version of the \eom\ (\ref{eomuetc}) on a periodic cartesian grid, using a 7-point stencil for the 3D-Laplacian and a leapfrog scheme for the time evolution. As mentioned above, we keep  a constant comoving scalar mass $m_c = \sqrt{\la_c}v = am_{\Phi}$. Space and time coordinates $x^\mu$ are measured in terms $m_c^{-1}$ units, whereas field scalar amplitudes in units of $v$, and field derivatives in units of $m_cv$.  The grid size $N^3$ of our simulations varies from $1024^3$ to $2048^3$, with lattice spacings $\De x = 0.5\,m_c^{-1}-1.0\,m_c^{-1}$ and timestep $\De t = 0.2\De x$.
 
In our simulations, we are not interested in the initial configuration of the fields; what we are after is the scaling regime. Once scaling is reached, the memory of the initial configuration is gone, so the only importance of the initial configuration is that they should lead the system to scaling as fast as possible. 
We initialize the fields with independent random values constrained to lie on the vacuum manifold, \ie~within a $(N-1)$-sphere $\Phi^{\rm T}\Phi = v^2/2$, and with $\dot\Phi = 0$.  After an initial transient time with diffusion evolution, we start evolving the fields with the lattice version of \eom\ (\ref{eomuetc})  (see Appendix~\ref{app:2}) and eventually the system relaxes into the scaling regime. In practice we enforce that the diffusion phase lasts for as long as it takes to reach the condition $(1/2-|\Phi|^2/v^2)^{1/2} \leq 0.01$, which takes typically a time of the order of $t_{0.01} \approx 30\,m_c^{-1}$ for $N > 3$ and $t_{0.01} \approx 50\,m_c^{-1}$ for $N=2,\ 3$. In order to determine the time when the network reaches scaling, $t_{\rm sca}$, we find the time at which the correlation-length estimator based on the energy-density of the system begins to be linearly proportional to time \cite{Lopez-Eiguren:2016jsy}. We also complement this method by tracking the overlapping of the energy-momentum correlators (see below). For most cases, \eg~in RD, the system reaches scaling at a time $t_{\rm sca} \approx 80\,m_c^{-1} + t_{0.01}$, except for $N=2$, where scaling is reached later, $t_{\rm sca} \approx 130\,m_c^{-1} + t_{0.01}$.
 
\subsection{Unequal time correlators (\uetcs)}
\label{subsec:UETC}

The energy-momentum tensor for the model is given by
\begin{equation}
T_{\mu\nu}(\bx,t) = 2\partial_\mu\Phi^{\rm T}\partial_\nu\Phi + g_{\mu\nu} \mathcal{L}(\Phi)\,.
\end{equation}
The method of calculating \uetc s from classical lattice field theory simulations is well-documented \cite{Spergel:1990ee,Durrer:1998rw,Bevis:2006mj,Bevis:2010gj}, and we will just briefly summarize it here.

The only \uetcs\ we are interested in this work are the tensor \uetcs\, since those are the ones contributing to  the the GW. 
Taking the spatial Fourier transform of $T_{ij}$, the two tensor polarizations ($A = 1,2$)  are given by
\begin{eqnarray}
S^T_A(\bk,t)=\sqrt{\frac{t}{2}} \sum_{i,j} M^A_{ij}T_{ij}(\bk,t) ,
\label{sa}
\end{eqnarray}
where the projectors $M^A_{ij}$ are defined as
\begin{eqnarray}
M^1_{ij} = \frac{1}{2}(e^1_ie^2_j + e^2_ie^1_j),\\
M^2_{ij} = \frac{1}{2}(e^1_ie^1_j - e^2_ie^2_j),
\end{eqnarray}
in the vector basis where $k_ie^A_ie^B_j = 0$ and $\delta_{ij}e^A_ie^B_j = \delta^{AB}$. They obey
\begin{eqnarray}
\sum_{A} M^A_{ij}M^A_{lm} = \La_{ij,lm}\,, 
\end{eqnarray}
with $\La_{ij,lm}$ the projector onto the TT-part of a tensor, defined in Eqs.~(\ref{lambda}). The UETC is obtained as
\begin{equation}
\uetcT(x_1,x_2 ) =\frac{1}{2} \sum_A \vev{S^T_A(\bk,t_1 ) S^T_A(\bk,t_2 )^*}, 
\label{average}
\end{equation}
where the average is taken over a set of numerical simulations and a shell in Fourier space.
Note that the UETC obeys the symmetry $\uetcT(x_1,x_2) = \uetcT(x_2,x_1)$.  

We only compute the corresponding \uetcs\, once the system has reached scaling. We construct them by multiplying the Fourier transforms of the TT-projected energy-momentum tensors at equally spaced times $t_{\rm ref}<t_1<t_{\rm end}$, $\Delta t=10$ for this work. We made a conservative choice for the UETC extraction initialization time: $t_{\rm ref} = 128$ for all cases except for $N=2$ where $t_{\rm ref} = 200$; whilst we respect the half-box light crossing time for the last UETC extraction, \ie\ $t_{\rm end} = L/2$, where $L=1024$ for $N=2, 3, 4$ and $8$, $L=768$ for $N=12$ and $L=512$ for $N=20$. In order to obtain the power spectra we average over a shell of width $\De k = 2\pi/L$, where $L$ is the side length of the simulation volume. 

\begin{figure}
\begin{center}
\includegraphics[width=0.49\textwidth]{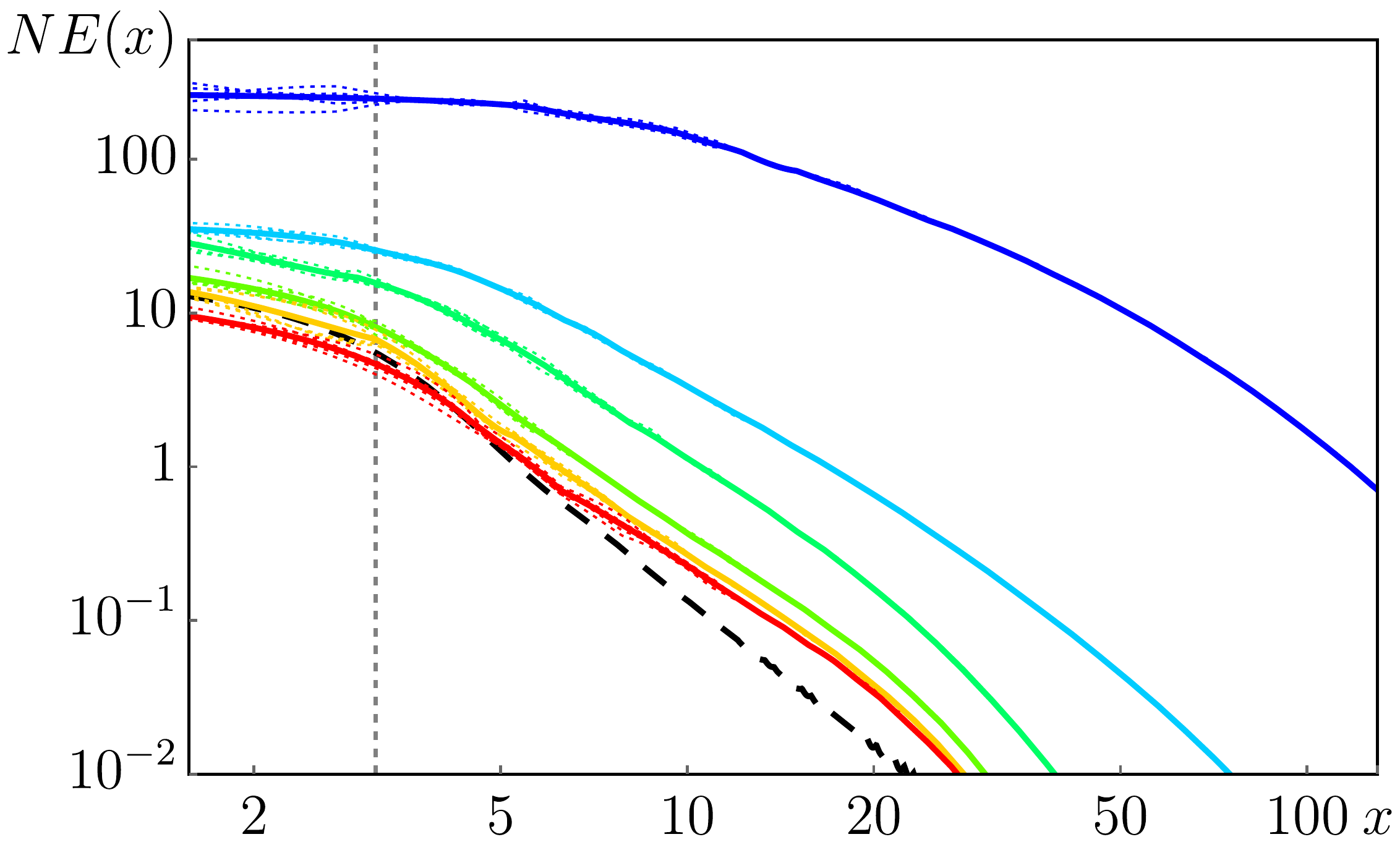}
\end{center}
\vspace*{-5mm}
\caption{ Comparison between the ETCs from numerical simulations $E_{\rm num}(x)$ (colour gradient) and theoretical analytical large-$N$ expressions (black dashed) $E_{\rm th}(x)$. All ETCs are obtained  at the reference time. The coloured lines correspond to (from top to bottom) $N$ = 2, 3, 4, 8, 12 and 20. All 5 realisations have been included, where each dotted line corresponds to individual runs and solids to the mean ETC. A dashed vertical line at $x=\pi$ is also included to show the point where $\Upsilon_{N}$ is computed.}
\label{fig:etcs}
\vspace*{-3mm}
\end{figure}

In Fig.~\ref{fig:etcs} we show the comparison between the ETCs from our numerical method above, denoted $E_{\rm num}(x)$,
with the one obtained in the large-$N$ limit analytical calculation, denoted $E_{\rm th}$. The numerically obtained ETCs  are the average over 5 different simulations. 
In the figure, we multiply the ETCs by $N$, as the output from the theory is the value of $NE_{\rm th}(x)$ as $N\to\infty$. We observe that the discrepancies are larger at higher $x=kt$ than at lower $x$, whereas the discrepancy is reduced the larger the value of $N$. We will therefore discuss the cases $N=2$ and $N=3$ separately.

 \begin{itemize}
 
 \item $N \geq 4$. The figure shows that  the  ETCs  for the $N=4, 8, 12$ and $20$ cases are close to the theoretical large-$N$ prediction. The larger the $N$, the closer the ETC approaches $E_{\rm th}$ (this effect is more evident at small $x$, i.e.~at larger length  scales).

\item $N = 2$ (strings) and $N = 3$ (monopoles). These two cases have the lowest value of $N$, and hence are the ones expected to be the furthest from the analytical prediction, because: 1) the field evolution is not linear, and 2) the vacuum manifold has non-trivial topology, and hence topological defects are expected to form. For the $O(2)$ case, the homotopy group of the vacuum is $\pi_1 \neq {\tt 1}$ and thus the defects are (global) cosmic strings,   
whose dynamics are rather different to the dynamics of 
non-topological field configurations. 
In particular, the time needed to reach scaling is larger, 
which makes the extraction of the scaling UETCs more complicated.
In the $O(3)$ case, the topology of the vacuum is also non-trivial  ($\pi_2 \neq {\tt 1}$), and global monopoles are formed in this model \cite{Bennett:1990xy,Achucarro:2000td,Lopez-Eiguren:2016jsy}. 
The departure from the large-N limit is however not as extreme as in $O(2)$. As we will see, the spectral amplitude of the GW background $\Omega_{\rm GW}$ deviates with respect the large-$N$ analytical computation by a factor $\mathcal{O}(100)$ and $\mathcal{O}(10)$, when evaluated at $N = 2$ and 3, respectively. 

\end{itemize}

In the coming section~\ref{subsec:GWfromUETC}, we will quantify the discrepancy of the numerical correlators against the analytical computation in the large-$N$ limit, and in particular how this impacts on the GW signal.

\subsection{Calculation of the GW signal}
\label{subsec:GWfromUETC}

As indicated by Eq.~(\ref{eq:GWspectrum3}), there is a direct connection between the GW energy density spectrum $\Omega_{\rm GW}$ and the UETC $\uetcT(x_1,x_2)$ for each value of $N$,
\begin{eqnarray}
 \Omega_{\rm GW} \propto \int dx_1 dx_2 \,{a_1a_2\over\sqrt{x_1x_2}}\,\cos(x_1-x_2)\,\uetcT(x_1,x_2),
\end{eqnarray}
where $a_1 \equiv a(x_1/k)$, $a_2 \equiv a(x_2/k)$. Numerically obtained UETCs from lattice simulations can be used to obtain the GW energy density power spectrum for each case, by a simple two-dimensional numerical integration. 
The specific formula for RD is given by Eq.~(\ref{eq:GWspectrum2}), whereas for MD is given by Eqs.~(\ref{eq:GWspectrum4}), (\ref{eq:F_MD}). In the following, we will use these formulas, and in particular their redshifted versions Eqs.~(\ref{eq:GWspectrumToday}) and (\ref{eq:GWspectrumTodayMD}), to calculate numerically the spectrum of the GWs emitted by a network of global defects in the O($N$) model. 
We will compare these numerical results (in particular for the RD case) with the analytical prediction based on the large-$N$ computation.

As observed in Sect.~\ref{subsec:UETC}, the smaller the $N$, the more the numerical UETC's deviate from the analytical prediction. A simple way to quantify this difference is to compare the equal time correlator (ETC) $E(x) = \uetcT(x,x)$ for each value $N$. In particular, we define 
\begin{eqnarray}
\Upsilon_N \equiv {E_{\rm num}(\pi) \over E_{\rm th}(\pi)}\,,
\end{eqnarray}
as the ratio between the numerical and theoretical computations of ETCs, evaluated at a scale $x = \pi$, corresponding to the moment when half wavelength has entered the horizon at RD.  The values we find for $\Upsilon$ are shown in Table \ref{tab:upsilon}. 
As expected, the numerical ETC approaches the theoretical prediction as $N$ grows, as indicated by the approach of $\Upsilon_N$ to unity as $N$ increases. For the case of cosmic strings ($N = 2$), the numerical ETC is a factor $\sim 100$ bigger than the analytical one, signalling a complete breakdown of the large $N$ approximation. 

We focus on the GWs produced during RD, as this is the relevant part of the spectrum for direct detection experiments. In order to obtain numerically the GW spectrum, we use the lattice version of Eq.~(\ref{eq:GWspectrumToday})
\begin{eqnarray}
h_0^2\Omega_{\rm GW}^{\rm num}\Big|_{\rm RD}  = h_0^2\Omega_{\rm rad}^{(0)}\left({v\over\Mpl}\right)^{\hspace*{-1mm}4}{\hspace*{-0.5mm}}{F}_{\rm RD}^{\rm (num)}\,,
\label{azkena}
\end{eqnarray}
with
\begin{eqnarray}
 {F}_{\rm RD}^{\rm (num)} &\equiv& {64\over3}\int_{x_{\rm min}}^{x_{\rm max}}\hspace*{-0.3cm} dx_1 \int_{x_{\rm min}}^{x_{\rm max}} \hspace*{-0.3cm} dx_2~ \sqrt{x_1x_2} \\
 && \hspace*{2cm}\times \cos(x_1-x_2)\,\uetcT(x_1,x_2)\,.\nonumber
\end{eqnarray}
The $\uetcT(x_1,x_2)$ are the numerical UETC's during RD, obtained for each value of $N$, and as explained before, ${F}_{\rm RD}^{\rm (num)}$ is the only quantity we need to extract from the simulations.
Even though in a lattice we are always bounded by an IR and UV scales (due to the finite volume and lattice spacing of the grid), we have made sure that ${F}_{\rm RD}^{\rm (num)}$ only changes marginally, whenever we change slightly the boundary values $x_{\rm min}$, $x_{\rm max}$, or by changing the number of points per dimension $N_p$. 
In all our simulations we see that  ${F}_{\rm RD}^{\rm (num)}$ converges rapidly and asymptotically for $x_{\rm max} \gg 1$ to the constant 
value ${F}_{\rm RD}^{(\infty)}$. Therefore, from our simulations we can  obtain ${F}_{\rm RD}^{\rm (num)} \approx {F}_{\rm RD}^{(\infty)}$.

All in all, we computed   $\uetcT(x_1,x_2)$  from our numerical simulations, which in turn can be turned into ${F}_{\rm RD}^{\rm (num)} \approx {F}_{\rm RD}^{(\infty)}$, and this can be substituted into Eq.~\ref{azkena} to obtain the numerical GW spectrum $\Omega_{\rm GW}^{\rm num}\Big|_{\rm RD}$.

In Table~\ref{tab:upsilon}, we also provide the lists of ratios of the asymptotic amplitudes of the GW spectrum during RD, comparing the lattice result to the theoretical amplitude obtained in the large $N$ limit, 
\begin{eqnarray}\label{eq:SigmaN}
\Sigma_N=\frac{\Omega_{\rm GW}^{\rm num}}{\Omega_{\rm GW}^{N\to\infty}}\,,
\end{eqnarray}
for different values of $N$.

\begin{table}[t]
Current data\vspace*{2mm}\\
\renewcommand{\arraystretch}{1.3}
\begin{tabular}{|| c | c c c c c c||}
\hline
    {${N}$} & {$ 2 $} & {$ 3 $} & {$ 4 $} & {$ 8 $} & {$ 12 $}  & {$ 20 $}  \\ \hline
$\Upsilon_N$ & 45 & 4.6 & 2.8 & 1.5 & 1.2 & 0.9 \\  \hline
$\Sigma_N$ & 238  & 10 & 4.1 & 1.7 & 1.3 & 1.0	\\ \hline
\end{tabular}\vspace*{5mm}\\
From {Paper I}\vspace*{1mm}\\
\renewcommand{\arraystretch}{1.3}
\begin{tabular}{|| c | c c c c c c||}
\hline
    {${N}$} & {$ 2 $} & {$ 3 $} & {$ 4 $} & {$ 8 $} & {$ 12 $}  & {$ 20 $}  \\ \hline
$\Upsilon_N$ & 36 & 4.5 & 3.1 & 1.7 & 1.4 & 1.3 \\ \hline
$\Sigma_N$ & 130 & 7.3 & 3.9 & 1.8 & 1.4 & 1.3\\ \hline
\end{tabular}
\caption{Values of the numerical ETCs at $x=\pi$, and GW amplitudes today, normalized to the large $N$ calculation. Top table: Ratios from simulations in this paper. Bottom table: ratios taken from Paper I. The statistical fluctuations are less than 10$\%$ in both cases. 
}\label{tab:upsilon}
\end{table}

If we fit the new numerical amplitudes against $N$, we obtain 
\begin{eqnarray}
\Omega_{\rm GW}^{\rm num}\Big|_{\rm RD} = \Omega_{\rm GW}^{N\to\infty}\Big|_{\rm RD}\left(a_0+  \frac{a_2}{N^2} + ...\right), 
\label{eq:NewApprox}
\end{eqnarray}
with $a_0=0.91 \pm 0.11$ and $a_2 = 51.1 \pm 3.5$, and negligible value for $a_1$. The above formula is valid only for $N \geq 4$.  The fit shows evidence that the numerical results converge to the large $N$ calculation as $N^{-2}$, albeit with a large coefficient,  confirming the result of Paper I.  

\begin{figure}[t]
\includegraphics[height=5cm]{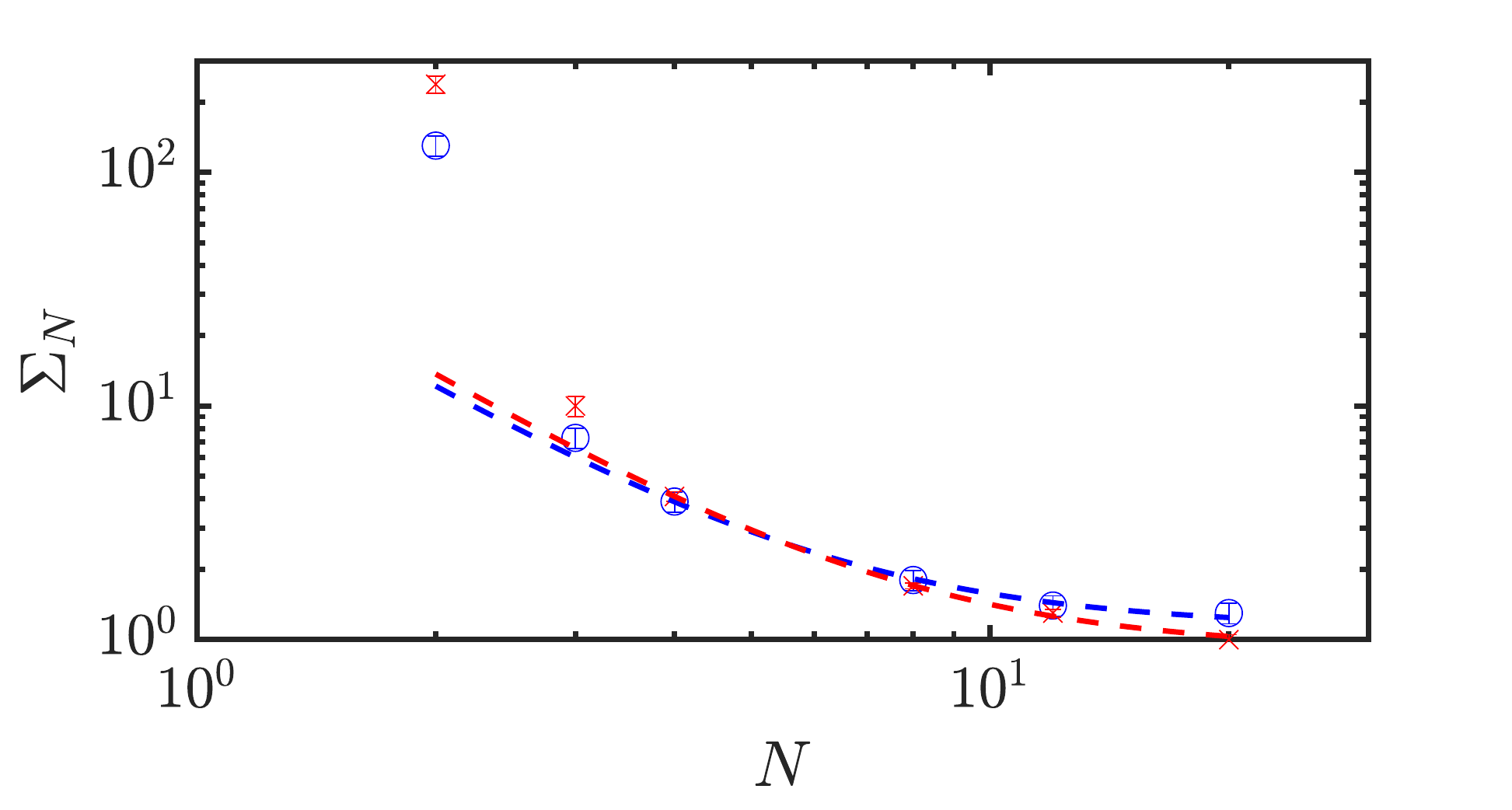}
\caption{$\Sigma_N$ from Table~\ref{tab:upsilon} and the comparison of the fit (\ref{eq:NewApprox}) for $N\geq4$ from paper I (blue circles) and current work (red crosses), including statistical uncertainties.. Data for $N=2$ and $N=3$ has also been included for completeness.}
\label{fig:FitSigmaN}
\end{figure}

The comparison between the results obtained in Paper I and the current work can be observed in Table~\ref{tab:upsilon}. Note that the statistical fluctuations between simulations are less than 10\%. Also, Fig.~\ref{fig:FitSigmaN} shows the comparison, where the standard deviations around the mean are depicted by the vertical lines. It can be seen that  the numbers obtained in Paper I and in the present work agree rather well and are consistent (except for the $N=2$ case, which we explain separately).

Turning to $N=2$, we can see in Table~\ref{tab:upsilon} that the value of $\Sigma_2$ has increased by roughly a factor of two. As mentioned earlier, this case is {special} because global cosmic strings are formed. It takes longer to reach scaling than for other cases, and therefore the time the network is simulated in scaling is shorter. Moreover, while the network length scale grows linearly in time, the intercept of the line with the $t$ axis is offset from zero, as explained in \cite{Bevis:2006mj,Bevis:2010gj,Daverio:2015nva,Lopez-Eiguren:2017dmc,Hindmarsh:2019csc}. 
This time offset is fed into the definition of the UETCs (note the factor of $\sqrt{t}$ in Eq. \ref{sa}), and therefore makes the UETC (and therefore the GW signal) larger. The fact that the value of $\Sigma_2$ has almost doubled can be accounted for by the value of the time offset in the simulations: whilst time offset was not considered for the values reported in Paper I, it is included in the computation of $\Sigma_2$ of this work.
Taking the this time offset into account in our old simulations reported in \cite{Figueroa:2012kw}, the numbers become closer. 
Some differences are also to expected because of the larger volume in this work.  
Further investigation is needed to understand and reduce the uncertainties in our measurement of $\Sigma_2$.

\subsection{Comparison with eigenvector decomposition.}
\label{subsec:GWfromDiagonalization}

A standard approach for computations of CMB fluctuations from topological defects is the decomposition of the 
{\uetc}s into a basis of its eigenvectors by diagonalisation, and then summing the power spectra resulting from each eigenvector, appropriately weighted by its eigenvalue.
This technique can also be applied to the GW power spectrum calculation.
In this section, we check the convergence of the partial sums over a series of weighted  eigenvector/eigenvalue terms to  
the GW energy density power spectrum obtained directly from the \uetc  .

\begin{figure}[t]
\includegraphics[width=8.7cm,height=5cm,angle=0]{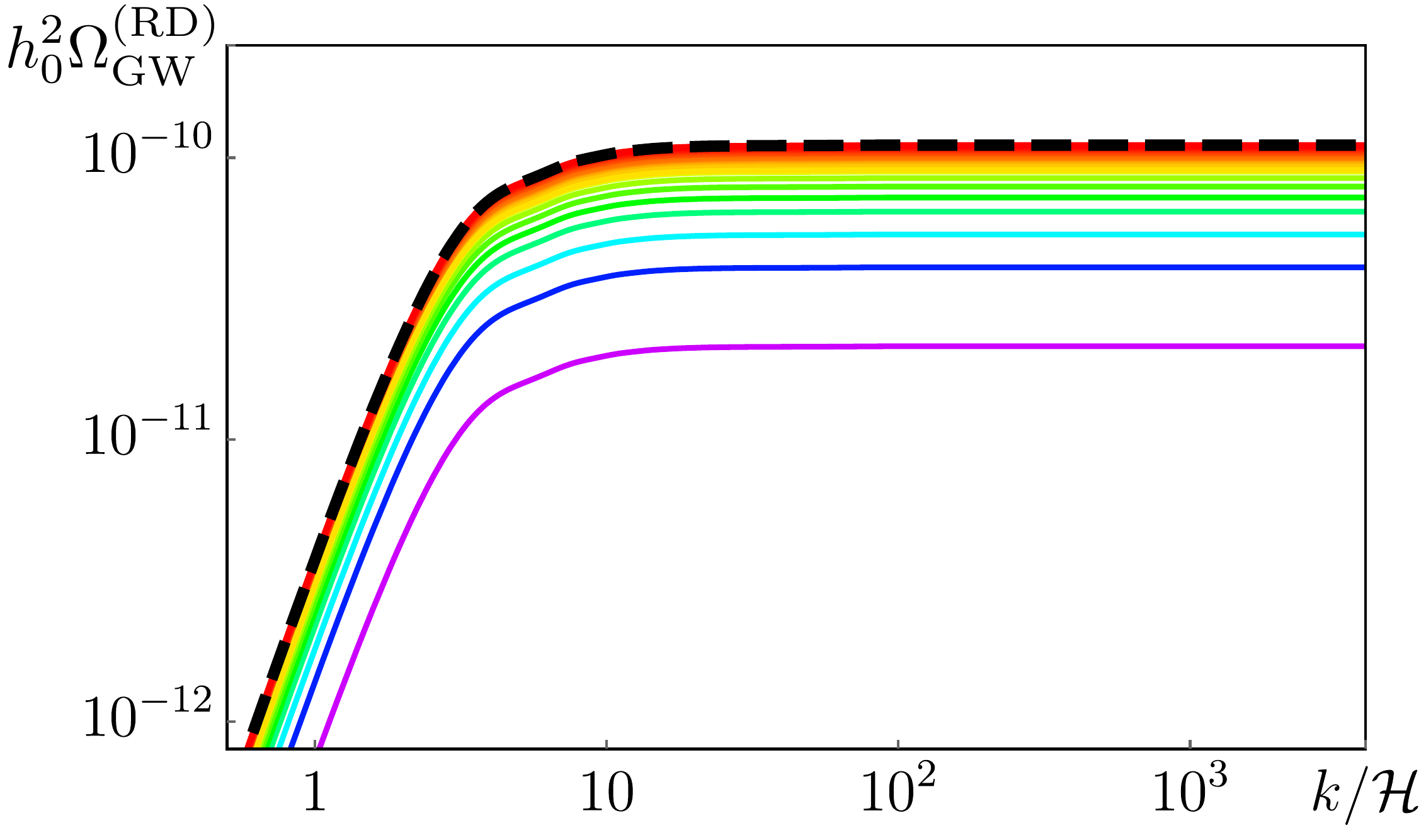}
\includegraphics[width=8.7cm,height=5cm,angle=0]{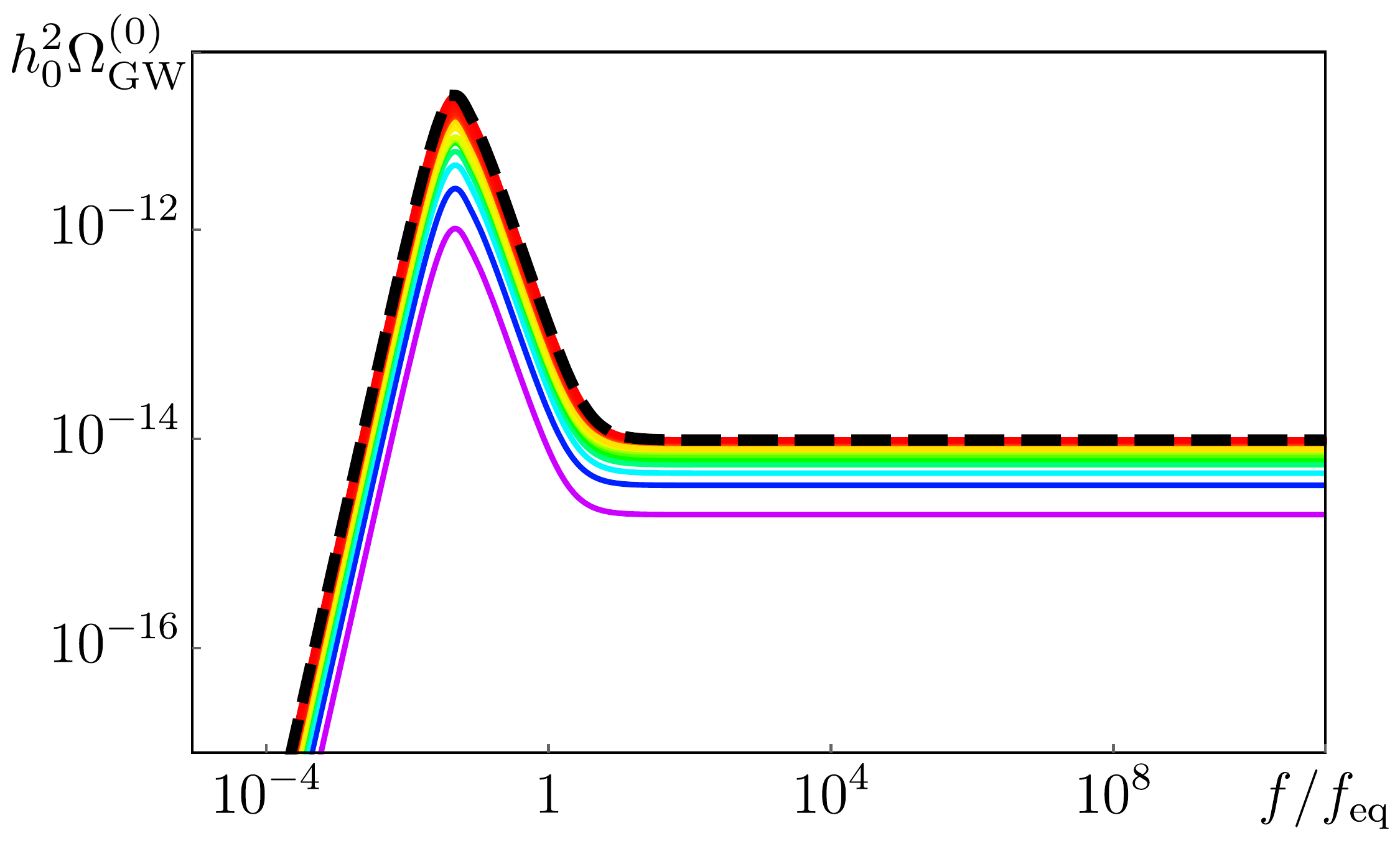}
\caption{Example of the GW background reconstruction from eigenvectors and eigenvalues in the $O(4)$ model during RD (top panel, with ${\mathcal{R}_*/ \mathcal{R}_t} = 1$) and today's full redshifted spectrum (lower panel, with ${\mathcal{R}_*} = 1$). The signals obtained from the full UETC is shown with black dashed lines.}
\label{fig:GWeigenValueVectorDecomp}
\end{figure}

Our \uetc\ is naturally discretized in $N_{s} = 2048$ steps in each each of the scaled wavenumber variables $x, x'$, so we are dealing with  $N_s \times N_s$ symmetric positive matrices. We can then diagonalize them, finding an orthonormal base of eigenvectors $\lbrace v_n(x)\rbrace$, with real positive eigenvalues $\lambda_i > 0$ that can be ordered as $\lambda_1 > \lambda_2 > \lambda_3 > ... 0$. Every UETC can then be written as
\be\label{eq:diag}
 \mathcal{U}(x_1,x_2) =\sum_n\la_nv_n(x_1)v_n^*(x_2) \,,
\ee
with the eigenvalues such that $0<\la_{n+1}<\la_n$. This can be applied to both UETCs from RD and MD. In the case of RD, Eq.~(\ref{eq:diag}) allows${F}_{\rm RD}^{\rm (num)}$ to be decomposed as 
\begin{eqnarray}
 {F}_{\rm RD}^{\rm (num)} = \sum_n  \la_{n}\left\lbrace\left|\mathcal{S}_{\rm RD}^{\rm (n)}\right|^2+\left|\mathcal{C}_{\rm RD}^{\rm (n)}\right|^2\right\rbrace\,,
 \end{eqnarray}
 with
\begin{eqnarray}\label{eq:Sn}
\mathcal{S}_{\rm RD}^{\rm (n)} \equiv {8\over\sqrt{3}}\int_{x_{\rm min}}^{x_{\rm max}}\hspace*{-0.3cm} dx \,x^{1/2} \sin(x)\,v_n(x)\,,\\
\label{eq:Cn}
\mathcal{C}_{\rm RD}^{\rm (n)} \equiv {8\over\sqrt{3}}\int_{x_{\rm min}}^{x_{\rm max}}\hspace*{-0.3cm} dx \, x^{1/2} \cos(x)\,v_n(x)\,.
\end{eqnarray}
This leads to 
\begin{eqnarray}
\Omega_{\rm GW}^{\rm num}\Big|_{\rm RD} &=& \sum_n \Omega_{\rm GW}^{\rm (n)}\Big|_{\rm RD}\\
 &\equiv& \sum_n \la_n \Omega_{\rm rad}^{(0)} \left({v\over\Mpl}\right)^{\hspace*{-1mm}4}{\hspace*{-0.5mm}}\left\lbrace\left|\mathcal{S}_{\rm RD}^{\rm (n)}\right|^2+\left|\mathcal{C}_{\rm RD}^{\rm (n)}\right|^2\right\rbrace\,.\nonumber
\end{eqnarray}
For MD, an equivalent expression can be written as
\begin{eqnarray}
&&\Omega_{\rm GW}^{\rm num}\Big|_{\rm MD} =
\sum_n \Omega_{\rm GW}^{\rm (n)}\Big|_{\rm MD}\\
 &&~~~~\equiv \sum_n \la_n \Omega_{\rm rad}^{(0)} \left({v\over\Mpl}\right)^{\hspace*{-1mm}4}{\hspace*{-0.5mm}}{k_{\rm eq}^2\over k^2}\left\lbrace\left|\mathcal{S}_{\rm MD}^{\rm (n)}\right|^2+\left|\mathcal{C}_{\rm MD}^{\rm (n)}\right|^2\right\rbrace\,,\nonumber
\end{eqnarray}
where $\mathcal{S}_{\rm MD}^{\rm (n)}, \mathcal{C}_{\rm MD}^{\rm (n)}$ are analogous expressions to Eqs.~(\ref{eq:Sn}), (\ref{eq:Cn}), but substituting ${8\over\sqrt{3}} \rightarrow {8\over\sqrt{3}}(\sqrt{2}-1)$ and $x^{1/2} \rightarrow x^{3/2}$ (inside the integrals).

In Fig.~\ref{fig:GWeigenValueVectorDecomp} we show the reconstruction of the GW spectrum for the $O(4)$ model during RD, as well as of today's full redshifted spectrum. Each line represents a spectrum reconstructed with the contribution of one more eigenfunction added; so the bottom line corresponds to having considered only the first term $n =1$, the next line above corresponds to having summed the first two terms $n = 1,2$, and so on so forth, all they way up to the highest line, which represents the sum of all the terms. The signal from direct integration of the \uetc\ is indicated with dashed lines.

As each term we add in the series is weighted by successively smaller eigenvalues, newer contributions contribute less and less. We observe that after adding only $\sim 15$ terms, the spectrum is already re-constructed to better than $~\sim 10\%$.

\section{Lattice computation of GWs. ~~~~ Part 2: Tensor real time evolution.}
\label{sec:GWfromRealTimeEvol}

Let us consider now the relativistic wave equation introduced in Section~\ref{sec:GW}, that governs the dynamics of GWs
\begin{eqnarray}
\ddot{{h}}_{ij}(\mathbf{x},t)+2\mathcal{H}\dot{h}_{ij}(\mathbf{x},t)-\nabla^{2}{h}_{ij}(\mathbf{x},t)={16\pi\over M_{\rm Pl}^2}\Pi_{ij}^{\rm TT}(\mathbf{x},t),\nonumber\\
\end{eqnarray}
with dots denoting derivatives with respect to the conformal time. In our case, the tranverse-traceless (TT) part of the anisotropic stress tensor $\Pi_{ij}^{\rm TT}$ -- the source  of the GWs --, is given by
\begin{eqnarray}
\Pi_{ij}^{\rm TT}(\mathbf{x},t) \equiv \sum_b(\partial_i\phi_b\partial_j\phi_b)^{\rm TT}\,,
\end{eqnarray}
with $\lbrace\phi_b\rbrace$ the $N$ componets of the scalar field field $\Phi = (\phi_1,\phi_2,...,\phi_N)^{\rm T}/\sqrt{2}$. For convenience, let us re-label the tensor perturbation as
\bea
{h}_{ij}(k,t) \equiv 16\pi\left({v\over M_{\rm Pl}}\right)^2 w_{ij}(k,t)\,,
\label{eq:NewHij}
\eea
so that their \eom\ can be written in terms of the dimensionless field variables ${\tilde \phi}_a \equiv \phi_c/v$, as 
\begin{eqnarray}\label{eq:wVariable}
\ddot{{w}}_{ij}+2\mathcal{H}\dot{w}_{ij}-\nabla^{2}{w}_{ij} =  \sum_b (\partial_b\tilde\phi\partial_b\tilde\phi_b)^{\rm TT}\,.
\end{eqnarray} 
The spectrum of the GW energy density contained within a volume $V$ (\ref{eq:GWrhoContSpectrum}) can be written in terms of the $w_{ij}$ variables as
\bea
\frac{d\rho_{\GW}}{d\log k}(k,t) = \frac{4k^3v^4}{\pi a^2(t)M_{\rm Pl}^2 L^3}{\left\langle \dot{w}_{ij}(k,t) \dot{w}_{ij}(k,t)\right\rangle_{\hspace*{-0.5mm}4\pi} }\,,\label{eq:GWspectrumNaturalTensors} 
\eea
where we have introduced $V = L^3$, with $L$ the length side of the lattice, and defined {\small$\left\langle ... \right\rangle_{\hspace*{-0.2mm}4\pi} \equiv {1\over 4\pi}\int\hspace*{-0.5mm} d\Omega_k\,...$}, with {\small$d \Omega_k$} a solid angle differential in {\bf k}-space.

In order to solve numerically the \eom\ for the GWs in the lattice, Eq.~(\ref{eq:wVariable}), we have followed the procedure originally introduced in~\cite{GarciaBellido:2007af}. We solve (a lattice version of) a relativistic wave equation for an unphysical perturbation $u_{ij}$ 

\be 
\ddot{{u}}_{ij}+2\mathcal{H}\dot{u}_{ij}-\nabla^{2}{u}_{ij} =  \sum_b (\partial_i\tilde\phi_b\,\partial_j\tilde\phi_b)\,, \label{eq:GW-nonTTeom}\ee
with no TT-projection over the source. 
We can then recover the physical TT part {\small$w_{ij}$} at any moment through 
\begin{eqnarray}
w_{ij}(k,t) = \Lambda_{ij,lm}(\hat k)u_{lm}(k,t)\, ,
\end{eqnarray}
with {\small$\Lambda_{ij,lm}(\hat k)$} the transverse-traceless projector Eq.~(\ref{projector}). Since {\small$\Lambda_{ij,pq}(\hat k)\Lambda_{pq,lm}(\hat k) = \Lambda_{ij,lm}(\hat k)$}, the argument inside the angular-average {\small$\langle ... \rangle_{\hspace*{-0.2mm}}$} in Eq.~(\ref{eq:GWspectrumNaturalTensors}), can be computed as
\begin{gather}
\dot{w}_{ij}(k,t)\dot{w}_{ij}(k,t) = 
\dot{u}_{ij}(k,t)\Lambda_{ij,lm}(\hat{k})\dot{u}_{lm}(k,t)\,.
\end{gather}
Appendix~\ref{app:3} explains this procedure of obtaining GW in a lattice.

We have studied the real time GW generation process for a model with $N=4$ scalar fields, in lattices up to {\small$N = 2048$} points per dimension. To solve the scalar field dynamics we have used the same standard lattice formulation as in Sect.~\ref{sec:GWfromUETC}.  In all simulations we have ensured that the lattice resolution covers well the dynamical range of momenta excited in the process, for both the scalar fields and the GWs, see e.g.~discussion in Sect.~\ref{subsec:LatticeParam}.

Defining $d{\tilde x} = m_{\rm c} dx$ as the dimensionless lattice spacing, with $m_{\rm c} \equiv \sqrt{\lambda_c }v$ and $\lambda_c = a^2\lambda = const$, so that $z = m_c t$ and $' \equiv {\partial/\partial z}$, the final expression of the GW spectrum in the lattice reads
\bea \label{eq:GWlattice}
\frac{d\rho_{\GW}}{d\log k}(\tilde{\bf n},z) &=& {4\over\pi}\left({v\over M_{\rm Pl}}\right)^4{d{\tilde x}^3\,\kappa({\bf \tilde{n}})^3\over N^3}{\lambda_c v^2M_{\rm Pl}^2\over a(z)^4}\\
&& \times ~a^2(z)~{\left\langle {u}_{ij}'(\tilde{\bf n},z)\Lambda_{ij,lm}^{\rm (L)}(\hat{k}){u}_{ij}'(\tilde{\bf n},z)\right\rangle_{\hspace*{-0.5mm}4\pi} }\,,\nonumber
\eea
with $\Lambda_{ij,lm}^{\rm (L)}(\hat{k})$ the lattice TT-projector, {\small$\kappa({\bf \tilde{n}}) \equiv k({\bf \tilde{n}}) / m_c$} the dimensionless lattice momenta, with {\small$k({\bf \tilde{n}}) \equiv k_{\rm IR}|{\bf \tilde{n}}|$} the momentum at the Fourier lattice site {\small${\bf \tilde{n}} = ({\tilde n}_1,{\tilde n}_2,{\tilde n}_2)$}, $-{N\over2} + 1 \leq {\tilde n}_j \leq {N\over2}$,  $k_{\rm IR} = {2\pi\over L}$ the minimum lattice momentum, and {\small$w_{ij} \equiv w_{ij}({\bf \tilde{n}},t)$} the discrete Fourier transform of {\small$w_{ij}({\bf n},z)$}, where {\small${\bf n} = (n_1,n_2,n_3)$} indicates the lattice sites, and $0 \leq {n}_j \leq N-1$.

In practice, we solve in the lattice the discretized version of the scalar fields' \eom~(\ref{eomuetc}) living in a background of RD with $a(z) \propto z$, together with Eq.~(\ref{eq:GW-nonTTeom}) for the (unphysical) spatial metric perturbations. We obtain, via Eq.~(\ref{eq:GWlattice}), the physical GW energy density spectrum at any time of the evolution. As explained at the beginning of Section~\ref{sec:GWfromUETC}, after setting up the initial random condition for the scalar field components, we evolve the system diffusively, until the scalar field expectation value reaches a small deviation (we choose $1\%$) with respect the true VEV, $\large|\sum_b {\tilde \phi}_b^2 - 1\large|^{1/2} \leq 0.01$, denoting this time by $z_{0.01}$. Once this condition is reached, the scalar field is allowed to follow its PRS-approximated equation of motion (\ref{eomuetc}). 

\subsection{Emergence of the GW plateau}
\label{subsec:Emergence}

\begin{figure}[t]
\includegraphics[width=8.7cm,height=6cm,angle=0]{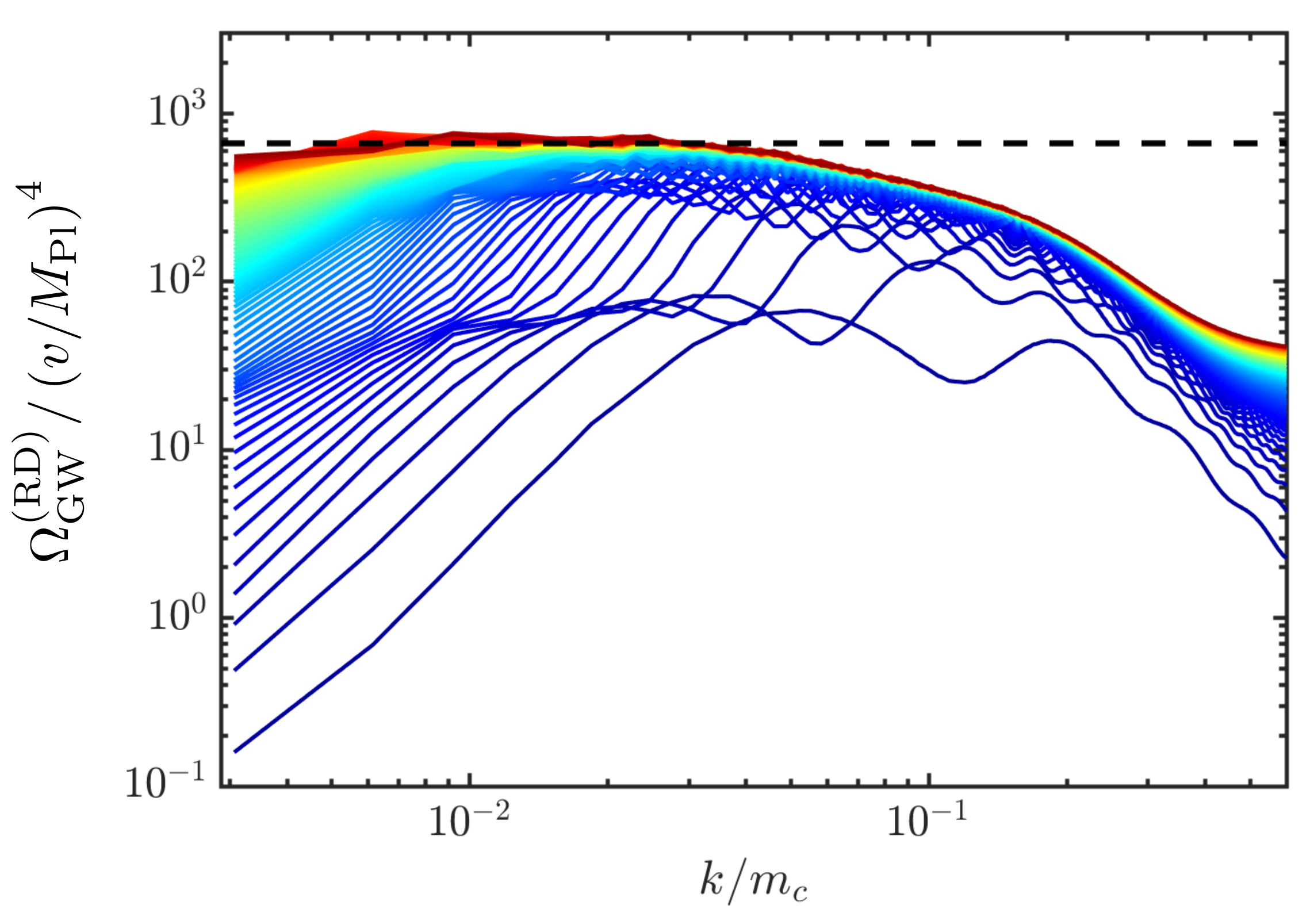}
\caption{Real time evolution of the GW energy density spectrum during RD, for $N_p = 2048, d\tilde x  = 1$, shown in time intervals $\Delta z = 20$, from $z_{\rm RD} = z_{\rm sca} = 100$ up to $z_{\rm final} = 2000$ (with ${\mathcal{R}_*/ \mathcal{R}_t} = 1$). At the most IR scales, from $\kappa = \kappa_{\rm IR} \simeq 0.3$ up to $\kappa_* \simeq 9k_{\rm IR} \simeq 3$, we see how the expected plateau is being gradually formed as time goes by. The plateau is actually well settled once the modes have entered well inside the horizon, at least an order of magnitude, see Fig.~\ref{fig:GWplateauRT2kdx1V2}. The final plateau settles down precisely at the same amplitude predicted by the UETC technique introduced in Sect.~\ref{sec:GWfromRealTimeEvol}, here indicated by dashed horizontal line.}
\label{fig:GWplateauRT2kdx1}
\end{figure}

We start the evolution of the \eom\ (\ref{eq:GW-nonTTeom}) for the GWs at a time $z_{\rm GW} \geq z_{0.01}$, choosing  initial condition $u_{ij}(\bx,z_{\rm GW}) = {\dot u}_{ij}(\bx,z_{\rm GW}) = 0$ at that moment. 
Solving the time evolution of the tensor perturbations leads then to a time dependent GW energy density spectrum, 
shown in Fig.~\ref{fig:GWplateauRT2kdx1} for a numerical simulation with $N_p = 2048$, and $d\tilde x = 1$.

Initially, the spectrum is peaked at a scale of the order $\sim \mathcal{O}(0.1) m_c^{-1}$. The amplitude of the GW spectrum then grows rapidly, and its shape changes, so that the initially suppressed IR part of the bump flattens out as time goes by. 
After some time, the GW energy density spectrum clearly exhibits a plateau in the IR, at least in the first nine wavenumber bins.

\begin{figure}[t]
\includegraphics[width=8.7cm,height=6cm,angle=0]{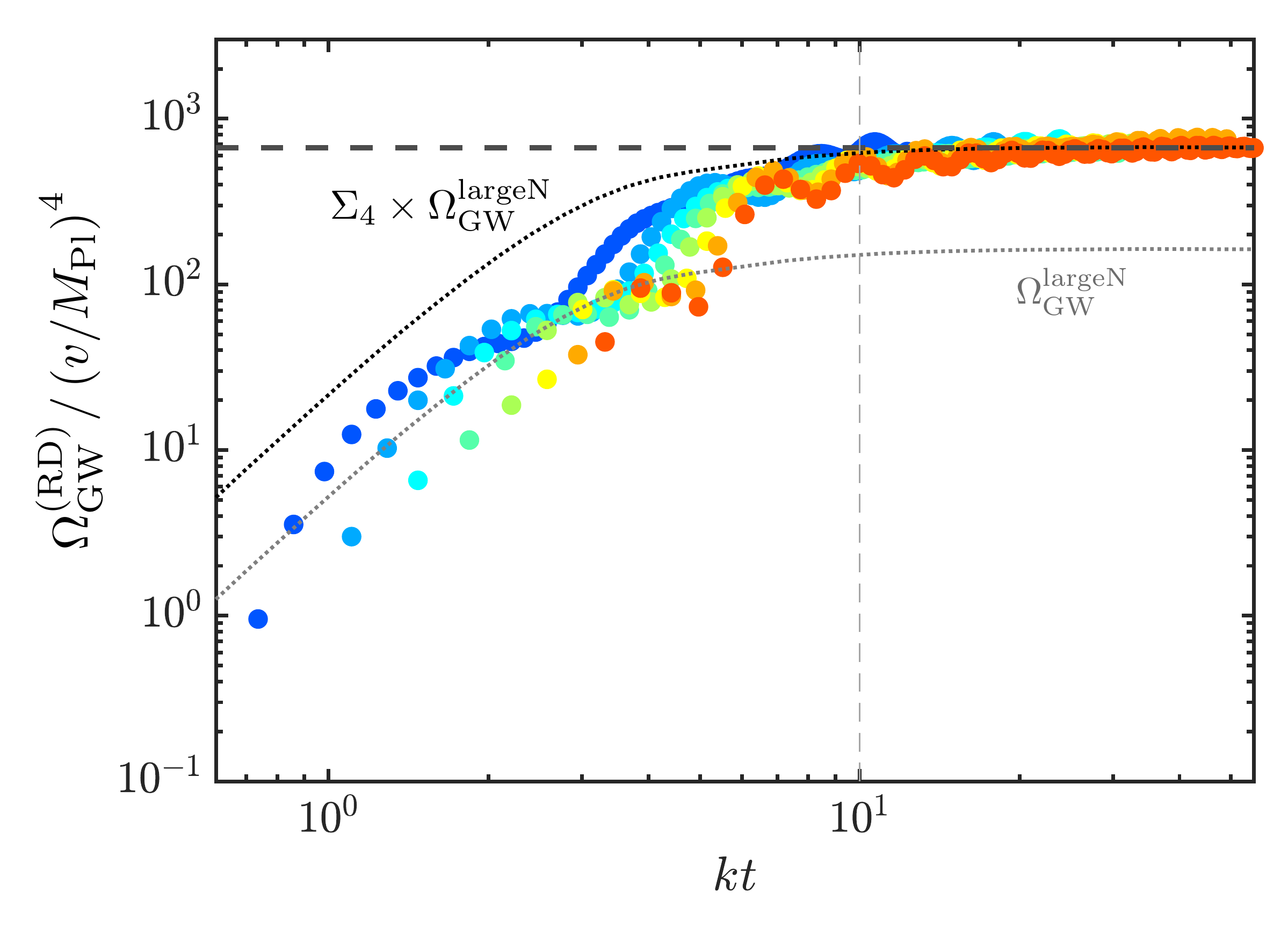}
\caption{The evolution of gravitational wave energy density in selected wavenumbers during RD, as a function of $x = kt$, for $N_p = 2048, d\tilde x  = 1$ (with ${\mathcal{R}_*/ \mathcal{R}_t} = 1$). We only plot the spectrum for the nine most IR modes of the simulation, which in Fig.~\ref{fig:GWplateauRT2kdx1} exhibit a plateau shape in the IR. Each mode is represented by a color, starting at blue (lowest $k's$), passing through green, yellow and orange, and ending in red (highest $k$ relaxing into the plateau). For comparison, we also plot the prediction from the large-N limit of the analytical calculation (lower gray dotted line), also re-scaled (higher black dotted line) by the compensating factor $\Sigma_4$ to match the UETC lattice results, c.f.~Eq.~(\ref{eq:SigmaN}).
}
\label{fig:GWplateauRT2kdx1V2}
\end{figure}

In Fig.~\ref{fig:GWplateauRT2kdx1V2}, we show the growth of $\Omega_{\rm GW}(k,t)$ for those wavenumbers $k$.
One sees that they saturate to a constant value after $k$ becomes sub-horizon, with , for $kt \gtrsim 10$. 
The most IR bin in our power spectrum does not have time to saturate, explaining why it is slightly below the plateau amplitude reached by the other modes.

Together, the graphs in Figs.~\ref{fig:GWplateauRT2kdx1} and~\ref{fig:GWplateauRT} show how the flat GW spectrum emerges on progressively larger scales, sourced on a scale close to the horizon scale. 
The growth, and therefore the sourcing, appears to stop for $kt \gtrsim 10$, which is presumably related to the field correlation length, slightly less than the horizon scale.
We interpret the scale of the initial bump as the initial correlation length of the field.

The behaviour of the modes outside the Hubble radius is a useful check of the numerical solution.  
From the analysis of the super-horizon Green's functions in Sect.~\ref{subsec:GWsubH}, the GW spectrum should exhibit an IR tail as $\propto x^3$, 
which implies a $k^3$ behaviour for the power spectrum, and a $t^3$ growth for the individual modes. 
Both these expected behaviours are visible in 
Figs.~\ref{fig:GWplateauRT2kdx1} and \ref{fig:GWplateauRT2kdx1V2}.

We now compare the GW spectra obtained from the real time evolution to that obtained from {\uetc}s in Sect.~\ref{sec:GWfromUETC}. 
The horizontal line in Fig.~\ref{fig:GWplateauRT2kdx1} represents the GW spectrum $\Omega_{\rm GW}(k,t)$ calculated via Eq.~(\ref{eq:GWspectrum2}) from the {\uetc}s. 
Both methods use data from the same simulations. 
The asymptotic plateau of the real-time GWs agrees very well with the amplitude of the GW spectrum derived from the {\uetc}s.

We also plot the large-N limit of the analytical calculation in Fig.~\ref{fig:GWplateauRT2kdx1V2}, depicted by the dotted gray line, which also exhibits the expected transition from $\propto x^3$ at large scales, to $\propto const.$ at smaller scales. If we re-scale such analytical prediction by the compensating factor $\Sigma_N$ obtained in Sect.~\ref{sec:GWfromUETC} for the $N=4$ model [based on the ratio of the GW spectra obtained from UETC's to the analytical large-N computation, c.f.~Eq.~(\ref{eq:SigmaN})], the amplitude of the re-scaled analytical prediction lies very close to 
the amplitude obtained for the plateau by the real time evolution.

The success of the previous comparisons between the UETC-based and real time GW spectra, provides a consistency check for both methods, and demonstrates that the use of either method should be considered equally acceptable in numerical computations of GWs from scaling seeds. This is one of the most important results of this paper.

\subsection{Importance of scaling for the GW source}
\label{subsec:SwitchON}

As the scalar field dynamics do not reach a {scaling} regime until $z = z_{\rm sca} > z_{0.01}$, a relevant aspect that needs to be quantified is the impact of different choices of $z_{\rm GW}$ in the GW dynamics.

First of all we should recall that the analytical predictions presented in Sect.~\ref{sec:GWfromCDmethods}, as well as the numerical computations based on the UETC's obtained in Sect.~\ref{sec:GWfromUETC}, are based on the scaling regime of the scalar field dynamics. If we switch on the GW evolution when the scalar field dynamics is not yet in scaling, i.e.~at some moment $z_{0.01} \leq z_{\rm GW} < z_{\rm sca}$, we expect the emerging GW spectrum to differ for each choice of $z_{\rm GW}$, as the GWs will experience different evolution histories which are not equivalent to each other by a simple `rescaling' of the size of the system into the horizon at each time. In other words, we expect that for GWs switched on too early, the resulting GW spectrum will be unphysical.

\begin{figure}[t]
\includegraphics[width=9cm,height=6cm,angle=0]{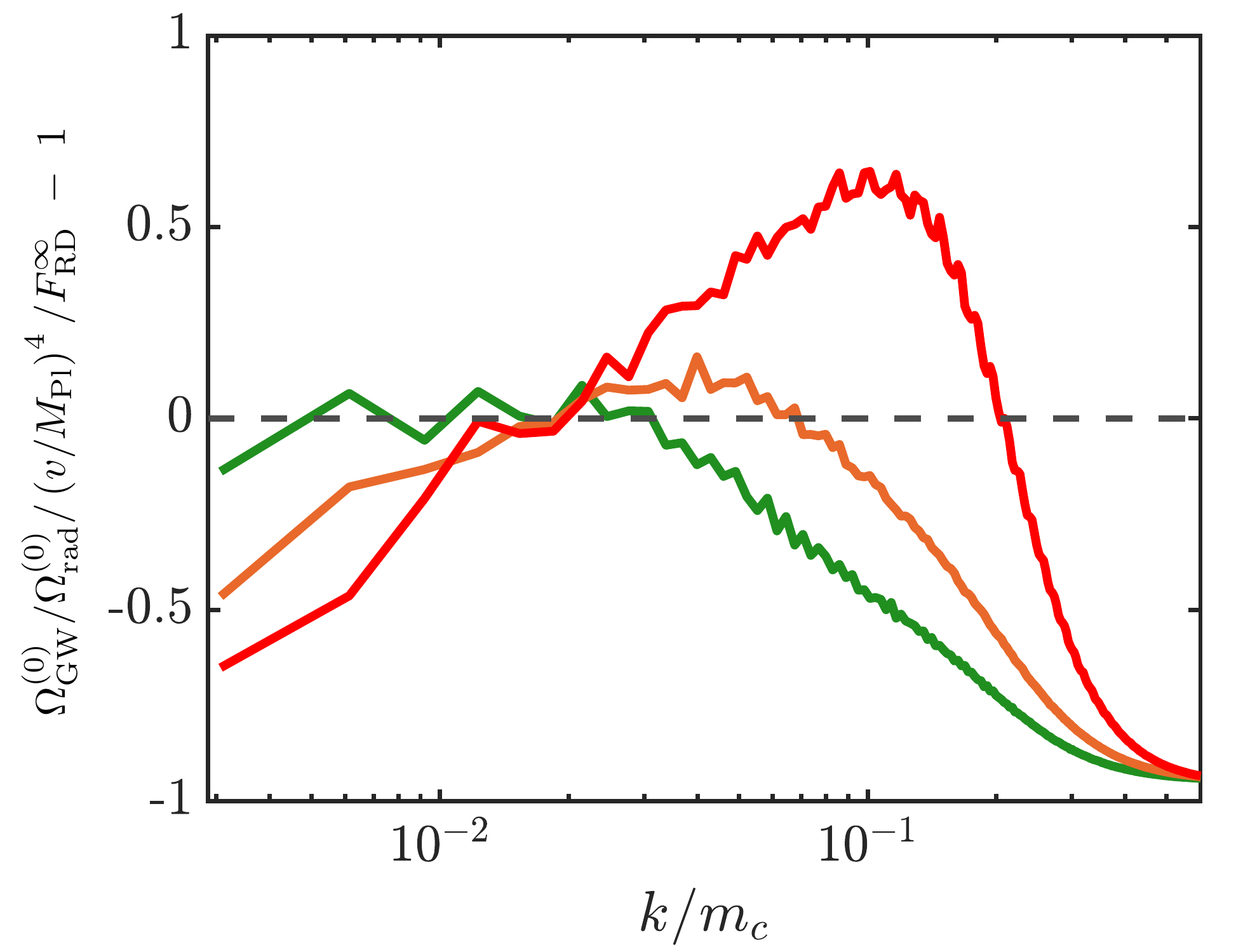}
\caption{Relative difference of the real time evolution GW power spectrum extracted at the end of each simulation (with ${\mathcal{R}_*/ \mathcal{R}_t} = 1$), with respect the plateau calculated from the UETCs $F_{\rm RD}^{(\infty)}$. All cases are obtained for $(N_p = 2048, d\tilde x=1.0)$, and correspond to $z^0_{\rm GW} = z_{\rm 0.01} = 50$ (red), $z_{\rm GW} = 75$ (orange), and $z_{\rm GW} = z _{\rm sca} = 100$ (green).}
\label{fig:GWplateauRT}
\end{figure}

In order to check the above phenomena, we performed simulations with several different values of $z_{\rm GW}$ in the range $z_{0.01} \leq z_{\rm GW} \leq z_{\rm sca}$. In Fig.~\ref{fig:GWplateauRT} we show the relative difference of GW spectra, extracted at the final simulation time, with respect the amplitude calculated from the UETC's, obtained when they are switched on at times $z_{\rm GW} = z_{0.01} + \Delta z_{\rm GW}$, with $\Delta z_{\rm GW} = 0$ (red) and $25$ (orange), together with the GW spectra for $z_{\rm GW} = z_{\rm sca}$ (green, which corresponds to $\Delta z_{\rm GW} = 50$). This figure clearly shows the importance of turning on the GW source when the network is already in scaling. An unphysical bump appears at scales $k/m_{\rm c}\sim 0.1$ when we start evolving GWs too early, which is nothing but the effect generated by the random initial conditions in the GWs energy density. The defect network is able to forget about the precise nature of its initial field configuration (as scaling implies), but its imprint in the GWs energy density spectrum remains and should be avoided. Furthermore, as the figure shows, turning the GW source too early also fails to create the IR plateau.  Hence the plateau is a feature of defect networks which emerges only once a scaling regime is sustained.

\subsection{Dependence on lattice parameters}
\label{subsec:LatticeParam}

As in any lattice simulation, we cannot choose arbitrary large volumes (i.e.~arbitrarily small $k_{\rm IR}$ scales), as the UV scales need also to be resolved with sufficient accuracy. We have therefore made sure that in our simulations the amplitude of  the GW spectra in the UV scales is well below the amplitude of the plateau in the IR scales (in some cases even exponentially suppressed when the UV coverage is good enough). The GW emitted at short wavelengths are related to the small scales in the problem, i.e.~to the characteristic microscopic scale of the defects $\gtrsim 1/m_c$. The dominant emission of GWs is rather expected due to the dynamics of the whole defect network, dictated by the scaling regime, and hence related to the horizon scale. As the defects self-order themselves around the horizon scale during {scaling}, GWs are emitted at the horizon scale at each moment of the evolution. We thus need to find a compromise between how well we can cover the IR scales (i.e.~how small $k_{\rm IR}$ can be), and how good we can resolve the microscopic scale $m_c^{-1}$ in the UV (i.e.~how large $dx$ can be tolerated, so that we still capture well the defect dynamics). In practice we find that a lattice spacing $d\tilde x \equiv m_c dx > 1$ leads to too large distortions of the UV part of the GW spectrum due to lattice artefacts in the defect dynamics, whereas a lattice spacing $d\tilde x < 0.5$ leads to a good exponential suppression of the UV tail of the GW spectrum, but only at the expense of 
the IR coverage, preventing the development of the IR plateau. In practice, we chose $d\tilde x = 0.5$ and $1.0$ and $N_p = 1024, 1512$ and $2048$. 

\begin{figure}[t]
\includegraphics[width=8.7cm,height=6cm,angle=0]{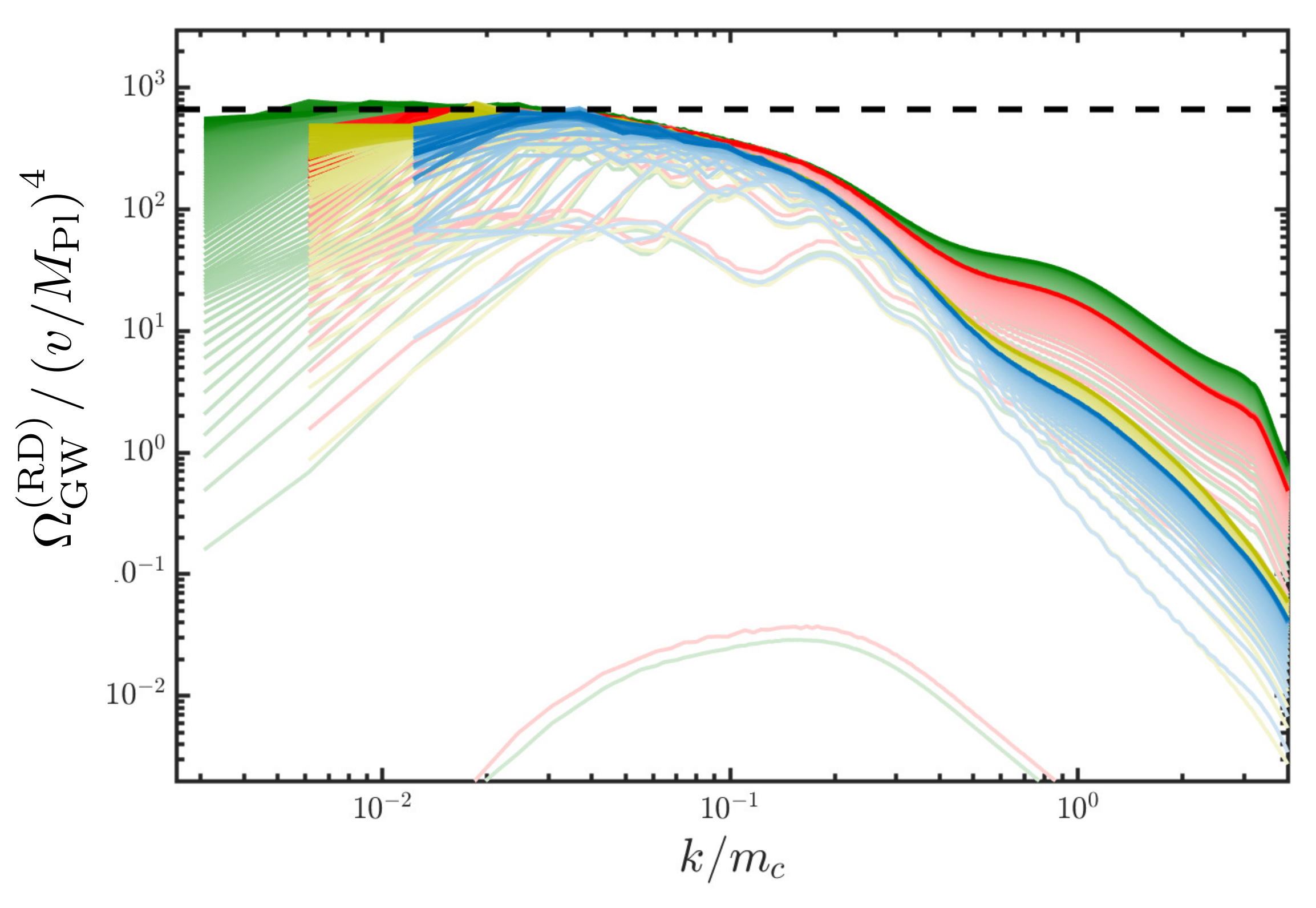}
\caption{Example of the real time evolution of the GW energy density spectrum (with ${\mathcal{R}_*/ \mathcal{R}_t} = 1$) for $(N_p = 1024, d\tilde x=0.5)$ [blue], $(N_p = 1024, d\tilde x=1.0)$ [red], $(N_p = 2048, d\tilde x=0.5)$ [yellow] and $(N_p = 2048, d\tilde x=1.0)$ [green].}
\label{fig:GWplateauRTV2}
\end{figure}

We have verified that simulations with different combinations of $\lbrace d\tilde x, N \rbrace$, lead to real time GW power spectra with a very similar amplitude in the overlapping IR region of wavenumbers. This can be seen in Fig.~\ref{fig:GWplateauRTV2}. Due to intrinsic limitations of our computer resources, the final IR plateau always spans over a finite range of momentum-scales  (in the best case scenario roughly around one decade, from the most IR scale $k_{\rm IR} = 2\pi/L$ of the lattice, up a scale $k \lesssim 10k_{\rm IR}$). The length of the plateau depends  on the lattice parameters $\lbrace d\tilde x,\tilde L \rbrace$. For the largest volume and lattice spacing that we have considered in a simulation ($N_p = 2048$, $d\tilde x = 1$), the presence of the IR plateau is clear for around one decade in wavenumbers, see e.g.~the green curve in Fig.~\ref{fig:GWplateauRTV2}. For smaller volumes ($N = 1024$) and/or lattice spacings ($d\tilde x = 0.5$), the plateau is also visible, but over a smaller range of wavenumbers. As shown in  Fig.~\ref{fig:GWplateauRTV2}, the plateau is always present in the overlapping range of wavenumbers shared by simulations with different values of $\lbrace dx,\tilde L \rbrace$. 
In particular, the plateau always appears at the same `turn-over' scale, $k_* \sim (0.03-0.04)m_c$, and spans smaller wavenumbers down to the characteristic $k_{\rm IR} = 2\pi/L$ of each lattice. 

In summary, the expected GW plateau always emerges during RD in all our simulations with different $N$ and $dx$. The length of the plateau varies however depending on the IR coverage of each simulation, and in the smaller volume simulations it is not readily apparent. 

\section{Summary and Discussion}\label{sec:Conclusions}

Cosmic defects are a natural by-product of a phase transition in the early Universe. The tensor metric perturbations they create 
are potentially observable as gravitational waves. The same tensor metric perturbations are also partly responsible for the B-mode polarization signal created by cosmic defects in the CMB. 

In this paper, we have calculated the GW spectra for defects from O($N$) global symmetry-breaking in two different ways: 
integrating the anisotropic stress unequal time correlator with the Green's functions for the tensor metric 
perturbations, and also by the real time simulation of the tensor perturbations sourced by the evolving defects. 
We find good agreement, demonstrating numerically the equivalence of the two methods, and providing 
a robust check on the results.  

Our results are consistent with, and improve on, those in Paper I \cite{Figueroa:2012kw}.  
The improvement is in two ways: we have extended the frequency range down to the Hubble rate today and beyond, 
giving a formal way to extend the power spectrum to super-Hubble scales.  
The numerical simulations are also twice as large, reducing uncertainties on the GW spectrum. 

The question arises then whether we could directly detect those tensor modes by GW experiments. 
Given the smallness of the frequency today corresponding to the horizon scale at the time of matter-radiation equality, c.f.~Eq.(\ref{eq:fEQ}),  direct GW detection experiments can only potentially probe the GW background produced during RD, corresponding to frequencies $f \gg f_{\rm eq} \simeq 6.6\cdot 10^{-17}$ Hz. Therefore, in order to assess the potential observability of the GW background from defects, only the GW plateau amplitude is relevant. Using Eqs.~(\ref{eq:GW_largeNfinal}) and (\ref{eq:650overN}), together with Eq.~(\ref{eq:SigmaN}), we obtain that the GW plateau amplitude today is
\begin{eqnarray}\label{eq:PlateauAmplitudeToday}
h_0^2\Omega_{\rm GW}^{(0)} &\simeq& h_0^2\Omega^{(0)}_{\rm rad} \times \Sigma_N \times {650\over N} \left(\frac{v}{\Mpl}\right)^{4}\nonumber \\
&\simeq& 2.63\cdot 10^{-15} \times {\Sigma_N\over N} \times \left({G\mu\over 10^{-6}}\right)^2\, ,
\end{eqnarray}
where we have used $h_0^2\Omega^{(0)}_{\rm rad} \simeq 4\cdot 10^{-5}$, and we have introduced the dimensionless parameter
\begin{eqnarray}
G\mu \equiv \pi \left(\frac{v}{\Mpl}\right)^{2}\,.
\end{eqnarray}
As the largest plateau amplitudes are obtained for the lowest values of $N$, we can focus only in the case of global cosmic strings ($N = 2$) and monopoles ($N = 3$). In each case $\mu$ has a different meaning, e.g.~for global strings it is the tension of the core of the string, whereas for global monopoles $\mu\delta$ is roughly the energy stored in the monopole core, where $\delta$ the width of the monopole. 

Based on the latest Planck results, CMB constraints on such global strings and monopoles,  lead to~\cite{Lopez-Eiguren:2017dmc} 
\begin{eqnarray}\label{eq:GmuCMBconstraints}
\begin{array}{rl}
10^{12}(G\mu)^2 < 0.031 &,\,N  = 2\vspace*{0.2cm}\\
10^{12}(G\mu)^2 < 0.73 &,\, N = 3\,.
\end{array}
\end{eqnarray}
In this work, we have quantified the ratio between the numerical and large-N analytical GW spectra, as
\begin{eqnarray}\label{eq:SigmaNoverN}
{\Sigma_N\over N} \simeq  
\left\lbrace\begin{array}{ll}
119 &,\,N  = 2\,,\vspace*{0.2cm}\\
3.33 &,\, N = 3\,,
\end{array}\right.
\end{eqnarray}
as shown in table~\ref{tab:upsilon}. Using Eqs.~(\ref{eq:GmuCMBconstraints}) and (\ref{eq:SigmaNoverN}), the  maximum amplitude of the GW plateau (\ref{eq:PlateauAmplitudeToday}) that we can obtain, is then
\begin{eqnarray}\label{eq:GWtodayAmpl}
\begin{array}{rl}
h_0^2\Omega_{\rm GW}^{(0)} < 9.7\cdot 10^{-15} &,\,N  = 2\,,\\
h_0^2\Omega_{\rm GW}^{(0)} < 6.4\cdot 10^{-15} &,\, N = 3\,.
\end{array}
\end{eqnarray}
These amplitudes are larger than the maximum amplitude expected (as bounded by current CMB constraints~\cite{Akrami:2018odb,Ade:2018gkx}) for the quasi-scale invariant GW background in slow-roll inflation~\cite{Caprini:2018mtu}, $h_0^2\Omega_{\rm GW}^{\rm (inf)} \lesssim 10^{-16}$. Amplitudes in Eq.~(\ref{eq:GWtodayAmpl}) are however too small to be observed by any planned direct GW detection experiment. For instance, based on the projected capabilities of LISA to detect a stochastic GW background~\cite{Caprini:2019pxz}, we conclude that the GW plateau for global strings cannot be detected by LISA with any significant signal-to-noise ratio, as this would required at least $h_0^2\Omega_{\rm GW}^{(0)} \gtrsim 10^{-13}$ at the LISA sensitivity peak of $f_p \sim 10^{-3}$ Hz. 
The proposed Big Bang Observer satellite mission, BBO, would be able to improve on the CMB limits, having 
a projected sensitivity to a cosmic background of $h_0^2\Omega_{\rm GW}^{(0)} \gtrsim 10^{-17}$ \cite{Corbin:2005ny}.
The Deci-hertz Interferometer Gravitational wave Observatory, DECIGO \cite{Seto:2001qf,Kawamura:2006up,Yagi:2011wg,Kawamura:2011zz,Sato:2017dkf}, with similar expected sensitivity, could possibly also detect the signal coming from global strings.

Our GW amplitude for $N = 2$ is almost twice as big as the result from Paper I, and 
should be taken only as a preliminary result, as discussed at the end of Sect.~\ref{subsec:GWfromUETC}. 
This is partly due to the fact that our present simulations have a larger volume (a factor 8 larger) than those in Paper~I, 
and partly due to a different method of extracting the scaling \uetc.
A detailed analysis and further simulations are under way.

The case $N = 2$ is of special physical interest, because global strings are 
an inevitable consequence of 
 global $U(1)$ symmetry-breaking after inflation, 
 and are therefore associated with axion-like particle (ALP) dark matter models. In recent years there has been a revived interest in studying such axion strings, see e.g.~\cite{Klaer:2017qhr, Saikawa:2017hiv, Vaquero:2018tib, Ferrer:2018uiu, Gorghetto:2018myk, Long:2019lwl,Buschmann:2019icd, Hindmarsh:2019csc}. 
 A GW signal from axion strings could be complementary to current detection strategies for axion-like particles. In the case of hidden axion sectors with no interaction with the SM (other than gravitational), it might represent, potentially, the only accessible signal. Given the relevance of this case, and the need to have under control the technical difficulties commented above, we plan to study the case of global strings in more detail elsewhere. 

Let us compare now our results to other  studies of GWs from global defects. Ref.~\cite{Kuroyanagi:2015esa} presents simulations of the real time evolution of tensors, similar to our simulations in Sect.~\ref{sec:GWfromRealTimeEvol}, but introducing an early MD epoch (due to quadratic inflaton oscillations during reheating) before the onset of RD. The numerical GW energy density spectrum they obtain exhibits, as expected, a high frequency tail $\Omega_{\rm GW} \propto 1/f^2$ for the modes that crossed the horizon during MD, as expected. The GW spectra also exhibits a bump at the IR/intermediate scales that cross the horizon during RD. As the background dynamics goes deeper and deeper into RD, the bump grows and seems to start flattening; a tail $\Omega_{\rm GW} \propto f^3$ is however always visible in the most IR scales captured in their simulations, see for instance the low frequency part of the GW spectra in Fig.~2 of~\cite{Kuroyanagi:2015esa}. Although there is no clear plateau, 
the authors interpret their peak power as an estimate of its value, while expressing caution that more dynamical range 
than afforded by their $N_p = 512$ points/dimension is needed.
Indeed, we found that $N_p = 2048$ points/dimension and large lattice spacing ($d\tilde x  =1$) were needed to show a
clear plateau developed in the IR, and even then it spanned no more than roughly one decade of scales (Figs.~\ref{fig:GWplateauRT2kdx1} and \ref{fig:GWplateauRTV2}). 

More recently, Ref.~\cite{Chang:2019mza} studied the GW emission from 
oscillating loops chopped off from the network of global strings, using a Nambu-Goto approximation to their dynamics 
(see also~\cite{Battye:1993jv}). Using the velocity-dependent one-scale (VOS) model for the string network evolution~\cite{Martins:1996jp,Martins:2000cs,Martins:2018dqg}, the authors conclude that the emission of GWs by oscillating loops can be significantly greater than the GW signal we obtained in the present work, 
estimating $h_0^2\Omega_{\rm GW}^{(0)} \sim 10^{-12}$ at LISA frequencies for $G\mu \sim 10^{-7}$, whereas 
we would predict $h_0^2\Omega_{\rm GW}^{(0)} \sim 10^{-14}$.

Their results are based on analytical studies expecting the loops to sustain both GW and Goldstone emissions~\cite{Vilenkin:2000jqa,Battye:1997jk}: if the Goldstone decay channel is slow enough, GWs will be emitted by each loop for as long as they remain existing, and hence a significant stochastic background of GWs will be built up from the contribution of all loops during their lifetime. While we have not performed a detailed analysis of the loops in our simulations, it is known that loops of Abelian Higgs strings produced in field theory network simulations decay much more rapidly that predicted in the Nambu-Goto approach~\cite{Hindmarsh:2017qff}, and this is very likely also to be true for global strings. Indeed, a recent dedicated lattice study to the decay of global string loops~\cite{Saurabh:2020pqe}, concludes that the global string loop lifetime is of the order the loop initial length $L$. This is in concordance with our numerical results, in the sense that in our GW computation we include the contribution from every possible field configuration in the string network, including that of loops, and we do not observe an accumulated emission of GW radiation from long lived loops. These aspects will require further investigation, particularly given the relevance of the case of global strings in relation to ALP dark matter models.

\acknowledgments

DGF (ORCID 0000-0002-4005-8915) is supported by a Ram\'on y Cajal contract by Spanish Ministry MINECO, with Ref. RYC-2017-23493. DGF acknowledges hospitality and support from KITP in Santa Barbara, where part of this work was completed. MH (ORCID ID 0000-0002-9307-437X) acknowledges support from the Science and Technology Facilities Council (grant number ST/L000504/1). JL (ORCID ID 0000-0002-1198-3191) and JU (ORCID ID 0000-0002-4221-2859) acknowledge support from Eusko Jaurlaritza (IT-979-16) and MCIU/AEI/FEDER grant Fondo Europeo de Desarrollo  Regional  (Grant  No.  PGC2018-094626-B-C21). This work has been possible thanks to the computing infrastructure of the ARINA cluster at the University of the Basque Country, UPV/EHU. This research was also supported in part by the National Science Foundation under Grant No. NSF PHY-1748958.

\vspace*{2cm}
\appendix
\section{Analytic solution in the large-$N$ limit.}
\label{app:1}

Let us consider a global theory where an $O(N)$ symmetry is spontaneously broken into $O(N-1)$. The starting point is an $N$-component scalar field $\Phi = (\phi_1,\phi_2,...,\phi_N)^{\rm T}/\sqrt{2}$ with lagrangian given by Eq.~(\ref{Lagon}). For $N \gg 1$, the dynamics of the Goldstone modes can be well described by a non-linear sigma model~\cite{Turok:1991qq,Boyanovsky:1999jg}, where we force the vacuum constraint $\sum_a\phi_a^2(\bx,t) = v^2$ by a Lagrange multiplier, with $v$ the vacuum expectation value (VEV). Normalizing the symmetry breaking field to its VEV , $\beta^a \equiv \phi^a/v$, each component of the field obeys the non-linear sigma model evolution equation
\begin{eqnarray}\label{e:sigma}
\Box\beta^a -(\partial_\mu\beta\cdot\partial^\mu\beta)\beta^a &=& 0  ~,
\end{eqnarray}
where $(\partial_\mu\beta\cdot\partial^\mu\beta) = \sum_a\eta^{\mu\nu}\partial_\mu\beta^a(\mathbf{x},t)
\partial_\nu\beta^a(\mathbf{x},t)$ and $|\beta(\bx,t)|^2 \equiv \sum_a\beta^a(\mathbf{x},t)
\beta^a(\mathbf{x},t) = 1$. In the large $N$-limit, we can replace the sum 
over components by an ensemble average $T(x) = \sum_a \eta^{\mu\nu}\partial_\mu\beta^a\partial_\nu\beta^a = 
N \langle \eta^{\mu\nu}\partial_\mu\beta^a\partial_\nu\beta^a \rangle = \bar T(t)$. By dimensional considerations, $T \propto \mathcal{H}^2$, or $\bar T(t) = T_0t^{-2}$, with $T_0>0$. Replacing the non-linearity in Eq.~(\ref{e:sigma}) by this expectation value, and Fourier transforming, we obtain a linear equation
\begin{eqnarray}\label{e:sigma2}
{\ddot\beta}_k^{a} + \frac{2\gamma}{t}{\dot\beta}_k^{a\,'}
  + \left(k^2-\frac{T_0}{t^2}\right)\beta_k^a = 0\,,
  \label{eomsigma}
\end{eqnarray}
where $\gamma \equiv d\log a/d\log\eta$, with \eg~$\gamma = 1$ for RD or $\gamma = 2$ for MD. The solution to Eq.~(\ref{e:sigma2}) for constant $\gamma$, and preserving the vacuum manifold constraint $|\beta(\bx,t)|^2 = 1$, is given by $T_0 = 3(\gamma +1/4)$ and
\bea \label{e:beta2}
\beta^a(\bk,t) =\sqrt{A}\left(\frac{t}{t_*}\right)^{3\over2}\hspace*{-1mm}
   \frac{J_\nu(kt)}{(kt)^\nu}\beta^a(\bk,t_*)\,, 
\eea
with $J_{\nu}(x)$ Bessel function of order $\nu \equiv \gamma + 1$, and $A \equiv 4\Ga(2\nu-1/2)\Ga(\nu-1/2)/3\Ga(\nu-1)$.

Here $\beta^a(k,t_*)$ is the $a$-th component of the field at the initial time $t_*$, and the normalization constant $A$ has been determined by imposing the condition $\langle|\beta(\bx,t_*)|^2\rangle = 1$ at the initial time. The analytical solution Eq.~(\ref{e:beta2}) shows explicitly that the self-ordering dynamics of the non-topological defects exhibits scaling. The condition $\beta^2 = 1$ actually introduces correlations between the different components of $\beta$, but these lead to corrections of order $1/N$~\cite{Jaffe:1993tt}, which in the large $N$-limit can be   neglected. On large scales, $\beta^2(\bx,t) \simeq \langle\beta^2(\bx,t)\rangle (1 +\mathcal{O}(1/N))$ is a very good approximation at all times $t \geq t_*$. See~\cite{Fenu:2009qf} for details.

In the limit $N \gg 1$, it is also possible to calculate analytically the GW power spectrum emitted by the evolution of the resulting self-ordering process of the non-topological textures. The GW amplitude for modes entering the horizon during RD was calculated in~\cite{Fenu:2009qf} as follows. Starting from Eq.~(\ref{eq:GW_spectra(Pi)}) we just need to calculate $\Pi^2(k,t,t')$ [see Eq.~(\ref{eq:UETC})] using the solution Eq.~(\ref{e:beta2}). As we only care about the field gradients as a source of GWs, the TT-part of the effective anisotropic stress tensor is $\Pi_{ij}^{\rm TT} = \left\lbrace\partial_\mu\phi^a\partial_\nu\phi^a\right\rbrace^{\rm TT}$, which in Fourier space reads
\begin{eqnarray}
\Pi_{ij}^{\rm TT}(\mathbf{k},t) = v^2\hspace*{-1.5mm}\int\hspace*{-1mm}\frac{d^3q}{(2\pi)^3}\,q_l\Lambda_{ij,lm}(\hat \bk)q_m\beta^a(\mathbf{q},t)\beta^a(\mathbf{k}\tiny{-}\mathbf{q},t)\,.\nn\\
\end{eqnarray}
The UETC  defined by Eq.~(\ref{eq:UETC}) can then be written as
\bea\label{eq:UETC4pointFunct}
&& \hspace*{-1cm}\left\langle {\Pi}_{ij}^\TT(\bk,t)\,{{\Pi}_{ij}^{\TT}}(\bk',t')\right\rangle \nn\\
&=& v^4\int \frac{d^3q\,d^3q'}{(2\pi)^6}\left( q^{\rm T}
\Lambda q\right)_{ij}\left({q'}^{\rm T}\Lambda q'\right)_{lm}\\
&& ~~\times\Big\langle\beta^a(\mathbf{q},t)
 \beta^a(\mathbf{k-q},t)\beta^{*b}(\mathbf{q}',t')\beta^{*b}(\mathbf{k-q},t')\Big\rangle\nn\\
&\equiv & (2\pi)^3\,{\Pi}^2(k,t,t')\,\delta^{(3)}(\bk-\bk'),\nn
\eea
where we used the notation $\left( q^{\rm T}\Lambda q\right)_{ij} \equiv q_l\Lambda_{ij,lm}q_m$. 

It can be shown~\cite{Jaffe:1993tt} that in the large N limit the field $\beta$ is Gaussian distributed initially, up to corrections $\sim 1/N$. As its time evolution is linear, $\beta$ will remain a Gaussian field, so we can determine higher order correlators via Wick's theorem. This is relevant in order to compute the UETC given  in Eq.~(\ref{eq:UETC4pointFunct}), characterized by the 4-point field correlator. By means of  Wick's theorem, we can reduce the 4-point function of the self-ordering fields into products of 2-point functions, $\langle\beta
\beta\beta\beta\rangle \sim \sum_{\rm pairs} \langle\beta \beta\rangle$. One obtains~\cite{Kunz:1997zs,Fenu:2009qf}
\begin{eqnarray}\label{eq:Pi2}
\Pi^2(k,t,t')= 
&& v^4\int \frac{d^3q}{(2\pi)^3} \,q^4
 \left[1-(\hat\bk\cdot\hat\bq)^2\right]^2\\
 &&\hspace*{0.8cm} \times~\mathcal{P}_\beta^{ab}(|\mathbf{q}|,t,t')
\mathcal{P}_\beta^{ab}(|\mathbf{k-q}|,t,t') \,,\nn
\end{eqnarray}
where
\begin{eqnarray}
\left\langle \beta^a(\mathbf{k},\eta)\beta^{*b}(\mathbf{k}',t')\right\rangle \equiv (2\pi)^3\mathcal{P}^{ab}_\beta(k,t,t')\de(\bk-\bk')\,.
\end{eqnarray}
If we assume that $\beta$ is initially aligned on scales smaller than the comoving horizon $~t_*$, and that it has an arbitrary orientation on scales larger than $t_*$, this corresponds to a white noise spectrum on large scales and vanishing power on small scales,
\be\label{e:coreta*}
\langle\beta^a(\bk,t_*)\beta^{*b}(\bk',t_*)\rangle = (2\pi)^3 6\pi^2t_*^3\frac{\de^{ab}}{N}\Theta(1-kt_*)\de(\bk-\bk')
\ee
where we have neglected the details of the decay of the correlator around $k\eta_* = 1$, and simply
modeled it with a Heaviside function. The amplitude $6\pi^2t_*^3$ at super-horizon scales is determined from imposing the condition $\beta^2(\bx,\eta_*) \simeq \langle\beta^2(\bx,\eta_*)\rangle$ everywhere in space (up to corrections of order $1/N$). Using Eqs.~(\ref{e:beta2}) and (\ref{e:coreta*}), we finally arrive at
\begin{eqnarray}
\mathcal{P}^{ab}_\beta(k,t,t') &=& \frac{\de_{ab}}{N} 6\pi^2A(tt')^{3/2}
\frac{J_\nu(kt)J_\nu(kt')}{(kt)^\nu(kt')^\nu}\nn\\
& \equiv & \frac{\de_{ab}}{N}f(k,t)f(k,t') ~, \label{eq:Phi2UETC}
\end{eqnarray}
with $f(k,t) \equiv \pi\sqrt{6A}k^{3/2}{J_\nu(kt)}(kt)^{3/2-\nu}$. Note that this corresponds to a totally coherent source, in the sense that its unequal time correlator $\mathcal{P}^{ab}_\beta(k,t,t')$ is a product of a function of $t$ and $t'$.

Combining Eqs.~(\ref{eq:GW_spectra(Pi)}), (\ref{eq:Pi2}), (\ref{eq:Phi2UETC}), we arrive at
\begin{eqnarray}\label{eq:GWlargeNanalytics}
\frac{d\rho_{\rm GW}(k,t)}{ d\log k} 
 = \frac{G\,v^4}{4\pi^4}\frac{k^3}{a^4(t)}
 \int dt' dt''a(t')a(t'')\int d\mathbf{p}~|\mathbf{p}|^4 \nn\\
\times \sin^4\theta\cos(k(t'-t''))\mathcal{P}_{\beta}^{ab}(p,t',t'')\,\mathcal{P}_{\beta}^{ab}(|\mathbf{k}-\mathbf{p}|,t',t'')\,.\nn\\
\end{eqnarray}
This formula is actually valid to describe the energy density spectra during either RD or MD, by simply choosing the appropriate scale factor behaviour and value of $\nu$ in Eq.~(\ref{eq:Phi2UETC}) ($\nu = 2$ for RD and $\nu = 3$ for MD). For instance, integrating numerically Eq.~(\ref{eq:GWlargeNanalytics}) with $\nu = 2$ and $a(t) \simeq \sqrt{\Omega_{\rm rad}^{(0)}}a_0^2H_0t$, one obtains the spectral plateau amplitude for the modes emitted during RD~\cite{Fenu:2009qf,DaniPhD}, that we reported in Eq.~(\ref{eq:650overN}).

\section{Lattice formulation of the scalar field equations}
\label{app:2}

The model under study is a model with  an $N-$component scalar field  $\Phi = (\phi_1,\phi_2,...,\phi_N)^{\rm T}/\sqrt{2}$ , with Lagrangian given in Eq~(\ref{Lagon}). This Lagrangian has an $O(N)$ symmetry that is spontaneously broken to  $O(N-1)$.  From this Lagrangian, one can obtain the corresponding \eom, which are the ones given in Eq.~(\ref{eomuetc}). 

In order to simulate the model in a lattice, we need the discrete version of those \eom. The approach we follow consists on the discretization of the action  \cite{Bevis:2006mj}, and from it we derive the \eom, rather than discretizing the \eom\ directly. Then, we use a leapfrog method to solve the equations in the lattice.

Before discretisation, we introduce the following dimensionless variables: $\phi_a \to \phi_a/v$, $dx \to m_{\rm c} dx$, with $m_{\rm c} \equiv \sqrt{\lambda_c }v$ and $\lambda_c = a^2\lambda = const$. With this changes, the discretized action reads
\bea
S_{\rm lat}&=&\Delta t \Delta x^3 \sum_{t,x}  \left[\frac{1}{2} a(t+\delta t) \sum_{a=1}^N \pi_a(\vec{x},t+\delta t)^2\right.\nonumber \\ &&-\half a(t)^2 \sum_{a=1}^N \sum_{i=1}^3  \left(\frac{\phi_a(\vec x+\delta \vec x_i,t)-\phi_a(\vec x,t)}{\Delta x}\right)\nonumber\\&&
\left.-\frac{1}{4} a(t)^4 \left(\sum_{a=1}^N  \left(\phi_a(\vec{x},t)\right)^2-1\right)^2\right]\,,
\eea
where $\delta \vec x_i$ denotes the three directions in the lattice, e.g., $\phi_a(\vec x+\delta \vec x_2,t)=\phi_a(x,y+\Delta x,z,t)$, $\delta t=\Delta t/2$, and 
we have defined the lattice conjugate momenta of the fields as:
\be
\pi_a(\vec{x},t+\delta t)=\frac{1}{\Delta t}\left(\phi_a(\vec{x},t+\Delta t )-\phi_a(\vec{x},t)\right)
\label{pi}
\ee
Note that while the field $\Phi$ lives in integer time steps, the conjugate momentum $\Pi=(\pi_1,\pi_2,...,\pi_N)$ lives in half-integer time steps, as needed for the 
leapfrog method, and the times where the scale factor $a$ is evaluated is chosen accordingly.

We can now obtain the discretized \eom\ from this discretized action:
\begin{widetext}
\bea
\pi_a(\vec{x},t+\delta t)=\left(\frac{a(t-\delta t)}{a(t+\delta t)}\right)^2 \pi_a(\vec{x},t-\delta t)+&\Delta t&
\left[\left(\frac{a(t)}{a(t+\delta t)}\right)^2\sum_{i=1}^3\frac{\phi_a(\vec x+\delta \vec x_i,t)+\phi_a(\vec x-\delta \vec x_i,t)-2\phi_a(\vec x,t)}{\Delta x^2}-\right.\nonumber\\&&\left.-\left(\frac{a(t)^2}{a(t+\delta t)}\right)^2 \left(\sum_{b=1}^N  \left(\phi_b(\vec{x},t)\right)^2-1\right)\phi_a(\vec{x},t)\right]
\eea
\end{widetext}
which together with the equation for $\phi_a(\vec{x},t+\Delta t )$ obtained from  (\ref{pi}):
\be
\phi_a(\vec{x},t+\Delta t ) = \phi_a(\vec{x},t) + \Delta t \pi_a(\vec{x},t+\delta t)
\ee
 are the equations used for updating the values of the fields as the dynamics progresses.

\section{GW on the lattice}
\label{app:3}

We now discuss the discretization to obtain the spectrum of GWs in the lattice, following Ref.~\cite{Figueroa:2011ye}. We have seen in the main text that the  spectrum of the energy density of a (statistically) homogeneous and isotropic GW background in the continuum is given by Eq.~(\ref{eq:GWrhoContSpectrum}), which we reproduce here for ease of reading:
\begin{equation}\label{eq:GWrhoContSpectrumbis}
\frac{d\rho_{\GW}}{d\log k} = 
\frac{k^3}{(4\pi)^3 G\,a^2(t)V} \int \frac{d\Omega_k}{4\pi}\,\dot h_{ij}(\bk,t)\dot h_{ij}^*(\bk,t)\
\end{equation}
where $d\Omega_k$ represents a solid angle element in $\bk$-space. The previous equation is valid in the limit of a very large volume $V$ encompassing all relevant wavelengths. The same condition in the lattice means that the volume $V = L^3= (Ndx)^3$ needs to encompass sufficiently well the characteristic wavelengths of the simulated GW background. 

In order to derive an analogous discrete expression to Eq.~(\ref{eq:GWrhoContSpectrumbis}) for a lattice of volume $V=N^3 dx^3$, we need first to specify our discrete Fourier transform (DFT) convention. We use
\begin{eqnarray}
&&f({\bf n}) = \frac{1}{N^3}\sum_{\tilde n}e^{-\frac{2\pi i}{N}\tilde{\bf n} {\bf n}}\,\tilde f(\tilde{\bf n})\,, \\ && \tilde f(\tilde{\bf n}) = \sum_{n}e^{+\frac{2\pi i}{N}\tilde{\bf n} {\bf n}} f({\bf n})\,,
\end{eqnarray}
where the index ${\bf n} = (n_1,n_2,n_3)$, with $n_i = 0,1,...,N-1$, labels our lattice sites in configuration space, whereas the index $\tilde{\bf n} = (\tilde n_1, \tilde n_2, \tilde n_3)$ labels the reciprocal lattice, with $\tilde n_i = -\frac{N}{2}+1, -\frac{N}{2}+2,...$ $-1,0,1,..., \frac{N}{2}$. Following  Ref.~\cite{Figueroa:2011ye}, one arrives at
\begin{eqnarray}\label{eq:GWrhoDiscreteSpectrum}
\left(\frac{d\rho_{GW}}{d\log k}\right)(\tilde{\bf n})&\equiv&  \frac{|k(\tilde{\bf n})|^3}{(4\pi)^3 G\,a^2(t)L^3} \cdot  \\ &\cdot&
\left\langle\left[dx^3\dot h_{ij}(|\tilde{\bf n}|,t)\right]\left[dx^3\dot h_{ij}(|\tilde{\bf n}|,t)\right]^*\right\rangle\,, \nonumber
\end{eqnarray} 
where $\left\langle\dot h_{ij}(|\tilde{\bf n}|,t)\dot h_{ij}^{*}(|\tilde{\bf n}|,t)\right\rangle$ is an average over configurations with lattice momenta $\tilde{\bf n}' \in [\,|\tilde{\bf n}|,|\tilde{\bf n}|+\delta\tilde n\,]$. The DFT becomes the continuous Fourier Transform  (CFT) in the continuum limit: $\,DFT\lbrace f({\bf n})dx^3 \rbrace \rightarrow CFT\lbrace f({\bf x}) \rbrace$. Therefore, the expression~(\ref{eq:GWrhoDiscreteSpectrum}) matches  the expression~(\ref{eq:GWrhoContSpectrumbis}) in the continuum limit. Expression~(\ref{eq:GWrhoDiscreteSpectrum}) highlights that the natural momenta in terms of which to express the lattice GW spectrum are the discretized version of the continuum ones ${\bf k} = \tilde{\bf n} k_{\rm IR}$, and not any of the lattice-momenta that one can define based on the choice of a lattice derivative.

As mentioned in the main text,  the TT metric perturbations follow equation (\ref{eq:h(Pi)}) in the continuum, where the TT part of a tensor is more easily obtained in Fourier space by means of  the projectors  $\Lambda_{ij,lm}(\hat\bk)$ given in Eqs~(\ref{projector}), which we also reproduce here: 
\begin{eqnarray}
&& \Lambda_{ij,lm}(\mathbf{\hat k}) \equiv P_{il}(\hat\bk)
P_{jm}(\hat\bk) - {1\over2} P_{ij}(\hat\bk)
P_{lm}(\hat\bk),\,\nonumber\\
&&P_{ij} = \delta_{ij} - \hat k_i \hat k_j\,,\hV \hat k_i = k_i/k\hI\hV \,.
\end{eqnarray}

On the lattice, one can construct a different projector for each different discretization of the spatial derivative, and in \cite{Figueroa:2011ye} it was shown that those different projectors give rise to differences in the very UV part of the GW spectrum. Nevertheless, for our case, those differences,  after integrating the spectrum over its Fourier modes, amount  only to a few per cent. Thus, all projectors are equivalent within those errors, and the projectors used in this work are the ones based on nearest-neighbour spatial derivatives \cite{Figueroa:2011ye}:
\begin{eqnarray}
&&\Lambda_{ij,lm}^{(L)}(\tilde{\bf n}) \equiv P^{(L)}_{il}(\tilde{\bf n})P^{(L)}_{jm}(\tilde{\bf n})-\frac{1}{2}P^{(L)}_{ij}(\tilde{\bf n})P^{(L)}_{lm}(\tilde{\bf n})\,,\nonumber\\
&&P^{(L)}_{ij}(\tilde{\bf n}) = \delta_{ij} - \frac{k^{(L)}_{i}k^{(L)}_{j}}{|k^{(L)}|^2}\,,\nonumber\\ &&k^{(L)}_{i}= 
2\frac{\sin(\pi \tilde{n}_i/N)}{dx}\,.
\end{eqnarray}

\bibliographystyle{h-physrev4}
\bibliography{BibAuto,BibManual}

\end{document}